# Calibrating Noise for Group Privacy in Subsampled Mechanisms

Yangfan Jiang
National University of Singapore
jyangfan@comp.nus.edu.sg

Xinjian Luo
National University of Singapore
xinjluo@comp.nus.edu.sg

Yin Yang
College of Sci. and Engr., Hamad Bin Khalifa University
yyang@hbku.edu.qa

Xiaokui Xiao
National University of Singapore
xkxiao@nus.edu.sg

## ABSTRACT

Given a group size $m$ and a sensitive dataset $D$, group privacy (GP) releases information about $D$ (e.g., weights of a neural network trained on $D$) with the guarantee that the adversary cannot infer with high confidence whether the underlying data is $D$ or a neighboring dataset $D'$ that differs from $D$ by $m$ records. GP generalizes the well-established notion of differential privacy (DP) for protecting individuals' privacy; in particular, when $m = 1$, GP reduces to DP. Compared to DP, GP is capable of protecting the sensitive *aggregate* information of a group of up to $m$ individuals, e.g., the average annual income among members of a yacht club. Despite its longstanding presence in the research literature and its promising applications, GP is often treated as an afterthought, with most approaches first developing a differential privacy (DP) mechanism and then using a generic conversion to adapt it for GP, treating the DP solution as a black box. As we point out in the paper, this methodology is suboptimal when the underlying DP solution involves subsampling, e.g., in the classic DP-SGD method for training deep learning models. In this case, the DP-to-GP conversion is overly pessimistic in its analysis, leading to high error and low utility in the published results under GP.

Motivated by this, we propose a novel analysis framework that provides tight privacy accounting for subsampled GP mechanisms. Instead of converting a black-box DP mechanism to GP, our solution carefully analyzes and utilizes the inherent randomness in subsampled mechanisms, leading to a substantially improved bound on the privacy loss with respect to GP. The proposed solution applies to a wide variety of foundational mechanisms with subsampling. Extensive experiments with real datasets demonstrate that compared to the baseline convert-from-blackbox-DP approach, our GP mechanisms achieve noise reductions of over an order of magnitude in several practical settings, including deep neural network training.

## 1 INTRODUCTION

With the rapid advances of machine learning techniques, data privacy has become a growing concern, and simple measures often fail to provide adequate protection to prevent leakage of sensitive information [26, 47]. Differential privacy (DP) [19, 20] is a strong and rigorous standard for ensuring individuals' privacy, which has gained adoption in industry [5, 15, 22] and widespread interest in academia [1, 8, 9, 16, 18, 33, 37, 40, 62]. In many practical scenarios, however, safeguarding only individual-level information may be insufficient, as aggregates over a group of individuals can also be highly sensitive [27, 28, 30, 31, 39, 49]. For instance, the income distribution of a bank's private banking customers can be a critical business secret. To tackle this issue, one natural approach is to extend the notion of DP to *group privacy* (*GP*), which protects the aggregate information of a group of individuals.

Specifically, a randomized algorithm $\mathcal{A}$ satisfies GP with group size $m$ if, for any pair of neighbor datasets $D$ and $D'$ differing by $m$ records, the output distributions of $\mathcal{A}(D)$ and $\mathcal{A}(D')$ are guaranteed to be indistinguishable in an information-theoretic sense, measured by specific privacy parameters, elaborated later in Section 2. In the special case that $m = 1$, this reduces to the classic DP definition. Note that the privacy guarantee in GP indicates that no group of $m$ records can have a significant impact on $\mathcal{A}$'s output distribution; hence, the larger the group size $m$, the stronger the guarantee. A foundational approach for achieving GP (which includes its special case DP) is to perturb the exact (i.e., non-private) result by injecting a calibrated amount of random noise [17]. Intuitively, a larger group size $m$, which corresponds to a stronger privacy guarantee, requires a higher amount of injected random noise to satisfy GP, leading to lower result utility.

Since DP is a hot research topic, many well-optimized solutions are available for enforcing DP in various problem settings. To satisfy the more general GP requirement, a common approach (e.g., [43]) is to convert an existing DP mechanism through a generic conversion procedure. Intuitively, since DP is a special case of GP with $m = 1$, we can obtain a GP-compliant mechanism by scaling the noise injected by DP by a factor that depends on the value of $m$. Note that here, the conversion algorithm is generic, meaning that it treats the underlying DP mechanism as a black box without considering the unique properties of the problem or the DP solution.

In this paper, we focus on the problem of releasing analysis results under GP, where the analysis involves random *subsampling*, as follows. Given an input dataset $D$ and an analysis algorithm $\mathcal{F}$, each record in $D$ is randomly selected into a subset $S$ with a probability of $q$ ($0 < q < 1$); after that, the analysis result is obtained by performing $\mathcal{F}$ on the subset $S$ instead of the entire dataset $D$. A notable example of subsampled analysis is stochastic gradient descent (SGD), which is commonly used to train large-scale machine learning models such as neural networks. It has been shown that for such applications, the amount of noise required to satisfy DP can be significantly reduced through sophisticated privacy accounting methods that exploit properties of subsampling [1, 6, 44, 53, 63], e.g., in the classic DP-SGD algorithm [1] for deep learning with DP. The intuition is that subsampling already provides a certain level of privacy protection, in the sense that as long as an individual record is not included in the sample set $S$, no information about this record is leaked even in the exact result $\mathcal{F}(S)$. This inherent privacy protection is amplified as the sampling rate $q$ becomes lower.

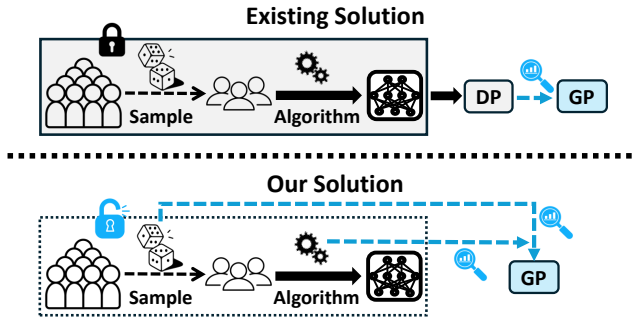

**Figure 1: High-level idea: the existing solution converts a black-box DP mechanism, while our approach conducts a direct, white-box-style privacy analysis.**

**Main observation.** To enforce GP with a given group size $m$ on subsampled analysis results, a baseline approach would be to take the corresponding DP mechanism and apply the above-mentioned generic conversion, by scaling the random noise by a factor determined by $m$. The problem with this approach is that the generic conversion, clarified in Section 2.2, treats the underlying DP mechanism as a black box and, thus, fails to utilize the inherent randomness introduced by subsampling, leading to suboptimal result utility. To illustrate, consider two neighbor datasets $D$ and $D'$ that differ by $m$ records, and a subsampling analysis that selects each record with probability $q$. Let $S$ and $S'$ be the sample sets obtained from $D$ and $D'$, respectively; for simplicity, let's consider the case that the same random seed is used to obtain both $S$ and $S'$. Then, intuitively, $S$ and $S'$ should differ by $qm$ records in expectation, rather than $m$, which has a vanishingly low probability of $q^m \ll 1$. To be more precise, the number of different records between $S$ and $S'$ follows the binomial distribution $B(m, q)$, with mean $qm$ and variance $q(1-q)m$. When $q$ is sufficiently low, e.g., $q = O(1/m)$, the variance of $B(m, q)$ becomes a constant, in which case the number of different records between $S$ and $S'$ is tightly concentrated around its mean value $qm$. This hints that we might be able to achieve GP by scaling the noise injected by DP by a factor that depends on $qm$, which is significantly smaller than $m$ as in the baseline approach.

**Our contributions.** In this paper, we establish a refined group privacy bound for subsampled mechanisms through a more sophisticated and precise privacy analysis. Our analysis follows the framework of Rényi group privacy (RGP) [43], which enables more accurate privacy guarantees for subsampled and iterative mechanisms such as DP-SGD [1]. Further, an RGP mechanism can be transformed to satisfy traditional notions of GP, as elaborated in Section 2.1. Unlike existing methods that simply convert a black-box DP mechanism, our approach is subsampling-aware, which directly analyzes the Rényi divergence between the output distributions of the subsampled analysis on pairs of neighbor datasets that differ by $m$ records, as illustrated in Figure 1. This direct, white-box-style analysis offers several advantages for tightening up the RGP guarantee: (i) it accounts for specific algorithmic characteristics, such as the shape of the output distribution; (ii) it harnesses the inherent randomness in subsampled mechanisms for amplifying the privacy guarantee; and (iii) it significantly reduces the impact of worst-case scenarios on group privacy guarantees. These properties help significantly enhance our RGP guarantee for subsampled mechanisms.

Through rigorous theoretical analysis, we prove that our bound, in general, offers a substantially improved RGP guarantee compared to previous methods for the subsampled mechanisms. For instance, in the case of the subsampled Gaussian mechanism [44], a core component in many widely-used privacy-preserving algorithms, including DP-SGD and its adaptations [3, 8, 27, 58, 63], our bound leads to noise reduction by a multiplicative factor of approximately $O(m^{0.58})$ compared to existing methods, where $m$ is the group size. Further, we prove the tightness of our general RGP bound for subsampled mechanisms. Specifically, we first establish an analytical lower bound of RGP guarantee for subsampled mechanisms by constructing a carefully crafted pair of neighboring datasets. Then, we show that our proposed RGP analysis asymptotically matches this lower bound, thereby justifying the tightness of our RGP bound.

Besides asymptotic improvements, the proposed RGP analysis framework has practical implications across various applications, in the sense that concrete instantiations of our RGP bound can be employed to derive significantly improved RGP guarantees for various privacy-preserving algorithms. Specifically, we present exact, *closed-form* RGP bounds for several widely-used mechanisms, including the subsampled Gaussian, Laplace, Skellam (which is often used to enforce DP in secure federated learning [3, 8]), and Randomized Response (commonly used in local DP [22]) mechanisms.

A result that might be of independent interest is that our analysis for the subsampled Laplace mechanism is not only a significant improvement for the RGP guarantee but for the popular Rényi differential privacy (RDP) definition as well. Specifically, for the $d$-dimensional Laplace mechanism, existing RDP methods, to our knowledge, all involve privacy composition, which accumulates privacy costs in each dimension. In other words, this approach incurs a multiplicative factor of $\Omega(d)$ in the privacy cost, which is prohibitively high when $d$ is large. To tackle this issue, we formulate the task of deriving a privacy guarantee for the $d$-dimensional Laplace mechanism as a constrained optimization problem, and demonstrate that this problem can be simplified to a more manageable one-dimensional scenario, thereby avoiding the composition.

Finally, we conduct a thorough comparison of the proposed GP bound with existing methods through numerical experiments. The results demonstrate a significant improvement of our method over existing ones, typically by over an order of magnitude in terms of injected error scale. Our results also validate the tightness of our RGP bound, which closely matches the theoretical lower bound under a wide range of configurations. In addition, we apply our results to enforcing GP on SGD for deep neural network training, using the MNIST, Fashion-MNIST, and CIFAR-10 benchmark datasets. The results show that compared to existing methods, our solution consistently achieves considerably higher model utility across different group sizes and privacy parameter settings.

## 2 PRELIMINARIES

### 2.1 Group Privacy

First, we formally define the notion of neighbor datasets, which is a major building block of the GP guarantee, as follows.

**Definition 2.1** (*m*-neighboring datasets). *Two datasets $D$ and $D'$ are m-neighboring datasets if and only if they differ by m records.*

In the above definition, parameter $m$ is referred to as the *group size*. When $m$ is clear from the context, we simply call $D$ and $D'$ neighbor datasets. Note that there are two cases that $D$ and $D'$ differ by $m$-records: (i) $D'$ can be obtained by adding or removing $m$ records from $D$, which is referred to as the *unbounded* definition [29], and (ii) $D'$ can be obtained by replacing $m$ existing records in $D$, which is called the *bounded* definition. Accordingly, GP defined using the bounded (resp., unbounded) definition of neighbor databases is referred to as bounded (resp, unbounded) GP.

Next we present the classic definition of GP, as follows.

**Definition 2.2** (($m, \epsilon, \delta$)-Group Privacy [21, 51]). *A randomized algorithm $\mathcal{A}$ satisfies ($m, \epsilon, \delta$)-DP if, for any two m-neighboring datasets $D, D'$, and for any subset of possible outputs $O \subseteq Range(\mathcal{A})$, it holds that*

$$\Pr\left[\mathcal{A}(D) \in O\right] \leq e^\epsilon \Pr\left[\mathcal{A}(D') \in O\right] + \delta. \tag{1}$$

A notable special case is that when $m = 1$, ($m, \epsilon, \delta$)-GP reduces to ($\epsilon, \delta$)-differential privacy [19, 20]. The parameters $\epsilon$ and $\delta$ control the trade-off between privacy and utility. Smaller values of $\epsilon$ and $\delta$ result in more similar output distributions of $\mathcal{A}$ on the neighbor datasets $D$ and $D'$, thereby providing stronger privacy protection. Meanwhile, as mentioned in Section 1, the group size $m$ also affects the GP guarantee: a larger $m$ prevents leakage of information (both individual and aggregate) derived from a larger group of individuals, which corresponds to a stronger guarantee, and vice versa.

Note that by definition, only a randomized algorithm can satisfy GP. Given a deterministic algorithm $\mathcal{F}$, we can ensure GP by injecting random noise to $\mathcal{F}$'s output [17], where the noise magnitude scales inversely with the privacy parameters $\epsilon$ and $\delta$, and at the same time positively correlated with the group size $m$ [20].

**Rényi group privacy (RGP) [43].** RGP is an alternative notion of group privacy that measures the indistinguishability between outputs of a randomized algorithm $\mathcal{A}$ using Rényi divergence [46]. RGP is particularly effective in accurately tracking cumulative privacy costs in iterative and subsampled mechanisms, due to its strong privacy composition property [43, 44], explained shortly. For this reason, the Rényi differential privacy (which is a special case of RGP with group size $m = 1$) plays a pivotal role in the analyses of the DP-SGD algorithm [1, 44] as well as its variants [3, 8, 27, 58, 63]. Formally, Rényi divergence and RGP are defined as follows.

**Definition 2.3** (Rényi Divergence [46, 52]). *Given two probability distributions $P$ and $Q$ that are defined on the same probability space $\mathcal{Z}$, let $P(z)$ and $Q(z)$ denote the densities of $P$ and $Q$ at $z \in \mathcal{Z}$. The Rényi divergence of a finite order $\alpha > 1$ is*

$$D_\alpha(P\|Q) = \frac{1}{\alpha - 1} \log \mathbb{E}_{z \sim Q}\left[\left(\frac{P(z)}{Q(z)}\right)^\alpha\right]. \tag{2}$$

**Definition 2.4** (($m, \alpha, \tau$)-Rényi Group Privacy). *A randomized algorithm $\mathcal{A}$ is said to satisfy ($m, \alpha, \tau$)-Rényi group privacy (RGP), if for any pair of m-neighboring datasets $D$ and $D'$, we have*

$$D_\alpha(\mathcal{A}(D)\|\mathcal{A}(D')) \leq \tau,$$

*where $D_\alpha(\cdot\|\cdot)$ is the Rényi divergence of order $\alpha$.*

When the group size $m = 1$, the ($m, \alpha, \tau$)-RGP is referred to as ($\alpha, \tau$)-Rényi differential privacy (RDP). The following lemma presents the sequential composition property of RGP. As mentioned above, this property is particularly useful for iterative mechanism design (e.g., in DP-SGD [1] and its variants), in which privacy cost accumulates across the iterations.

**Lemma 2.1** (Sequential Composition of RGP [43]). *Let $\mathcal{M}_1 : \mathcal{D} \mapsto \mathcal{R}_1$ and $\mathcal{M}_2 : \mathcal{R}_1 \times \mathcal{D} \mapsto \mathcal{R}_2$ be two randomized mechanisms with independent source of randomness that satisfy ($m, \alpha, \tau_1$) and ($m, \alpha, \tau_2$)-RGP, respectively. Then the combination of these two mechanisms, defined by $\mathcal{M}_{1,2}(D) := (\mathcal{M}_1(D), \mathcal{M}_2(\mathcal{M}_1(D), D))$, satisfies ($m, \alpha, \tau_1 + \tau_2$)-RGP.*

Finally, an ($m, \alpha, \tau$)-RGP can be converted into an ($m, \epsilon, \delta$)-GP guarantee for any $\delta \in (0, 1)$ through the following lemma:

**Lemma 2.2** (From RGP to GP [7, 13]). *If a randomized algorithm $\mathcal{A}$ satisfies ($m, \alpha, \tau$)-RGP, then for all $\delta \in (0, 1)$, the algorithm $\mathcal{A}$ also satisfies ($m, \epsilon, \delta$)-GP, where*

$$\epsilon = \tau + \frac{\log(1/\delta) + (\alpha - 1)\log(1 - 1/\alpha) - \log \alpha}{\alpha - 1}.$$

In practice, given a $\delta$ and an $m$, we determine the final ($m, \epsilon, \delta$)-group privacy guarantee by minimizing $\epsilon$ over $\alpha$ using Lemma 2.2. Typically, $\alpha$ ranges from 2 to 100, as commonly seen in production-ready libraries such as TensorFlow Privacy[1], Opacus[2], and autodp[3].

### 2.2 From RDP to RGP

We now introduce the existing general methodology for converting RDP guarantees to RGP guarantees.

**Lemma 2.3** (From RDP to RGP [43]). *Let $c \in \mathbb{N}$ be an arbitrary positive integer. If $\mathcal{A}$ satisfies ($\alpha, \tau$)-RDP and $\alpha \geq 2^{c+1}$, then*

$$D_{\alpha/2^c}(D\|D') \leq 3^c \tau,$$

*for all pairs of $2^c$-neighboring datasets $D$ and $D'$.*

Accordingly, to convert a ($\alpha, \tau$)-RDP to RGP with a given group size $m$, we can first find $c \in \mathbb{N}$ such that $2^c \geq m$, and then scale down $\alpha$ to obtain $\alpha' = \frac{\alpha}{2^c}$, and at the same time scale up $\tau$ to $\tau' = 3^c \tau$, to obtain the corresponding ($m, \alpha', \tau'$)-RGP guarantee. Observe that since the above lemma is only applicable for $c \in \mathbb{N}$, the conversion needs to increase $m$ to the next power of two. Since

---

$\tau$ is scaled up by a factor of $3^c$ which is approximately $m^{1.58}$ (since $\frac{\log(3)}{\log(2)} \approx 1.58$), this leads to a rather conservative bound on privacy cost, and, thus, high error scale required to satisfy RGP. As we shall elaborate in Sections 3 and 4, our results can derive $(m, \alpha, \tau)$-RGP guarantees for any $m \in \mathbb{N}$, overcoming this issue.

Lastly, we mention that references [21, 51] in which GP is defined also include a basic DP-to-GP conversion method that transforms a $(\epsilon', \delta')$-DP guarantee to $(m, \epsilon, \delta)$-GP, where $\epsilon = m\epsilon'$ and $\delta = \frac{e^{m\epsilon'}-1}{e^{\epsilon'}-1}\delta'$. Note that as $\epsilon'$ grows, $\delta$ approaches $e^m \delta'$, which can become excessively large even with a moderate group size $m$, e.g., 64. In practice, it is often required that $\delta$ is no more than $o(1/n)$ where $n$ is the number of records in the underlying dataset, since otherwise, one can release the exact value of a random record while still satisfying GP. Further, as mentioned before, the $(m, \epsilon, \delta)$-GP definition does not have the nice composition properties of RGP (i.e., Lemma 2.1) and, thus, is not easy to use for iterative algorithms such as SGD. Hence, this paper focuses on the RGP definition [43] and the corresponding RDP-to-RGP conversion rule described above.

### 2.3 Subsampled Mechanism

Let $D$ be a dataset and let $\mathcal{A}$ be a randomized algorithm. We adopt the definition of the subsampled mechanism used in the DP-SGD algorithm [1, 44], as follows.

**Definition 2.5** (Subsampled Mechanism). *Given an input dataset $D$, the subsampled mechanism constructs a subset $S \subseteq D$ by including each record $x \in D$ into $S$ independently with a fixed probability of $q \in (0, 1)$. $\mathcal{A}$ is then performed on $S$ to produce the privatized output. Formally, this is defined as:*

$$\mathcal{M}(D) \triangleq (\mathcal{A} \circ \text{Subsample})(D) = \mathcal{A}(\text{Subsample}(D)),$$

*where $\text{Subsample}(D)$ denotes the subsampling procedure that constructs the subset $S$ from $D$.*

The output distribution of $\mathcal{M}$ is essentially a mixture distribution, where the distributions of $\mathcal{A}(S)$ for all $S \subseteq D$ are the mixed components, and their corresponding mixture weights are the probabilities of $S$ being sampled. In other words, the distribution of $\mathcal{M}(D)$ can be expressed as:

$$\mathcal{M}(D) \sim \sum_{S \subseteq D} p_S \mathcal{A}(S),$$

where $p_S$ denotes the probability of constructing the subset $S \subseteq D$. Note that in the literature, different mechanisms may use different subsampling procedures. For instance, in DPIS [55], each record is selected into the sample set $S$ based on a probability computed based on an importance measure of the record. Another approach [53] involves uniformly sampling a subset from all possible subsets of a predetermined set size. In this paper, we focus on the most widely-used subsampling procedure, which samples each record independently with a constant probability $q$. This scheme is used in DP-SGD [1] as well as several of its variants [3, 8, 27, 58, 63]. The analysis of group privacy guarantees on other types of subsampled mechanisms will be considered in future work.

### 3 MAIN RESULTS

This section presents the main results of the paper: a refined, generic privacy cost upper bound for subsampled mechanisms under RGP.

This is essentially a subsampling-aware privacy accounting framework, whose concrete instantiations for specific subsampling mechanisms are presented later in Section 4. The key insight that we utilize for deriving the refined RGP bound is that the subsampling procedure amplifies group privacy. More detailed explanations of this insight behind our analysis are provided in Appendix A.

In what follows, we first introduce this generic RGP bound for subsampled mechanisms in Section 3.1 and derive its corresponding proof in Section 3.2. Then, in Section 3.3, we prove the tightness of our RGP bound.

### 3.1 General RGP Bound

The following theorem presents our general upper bound on the privacy cost under RGP for subsampled mechanisms.

**Theorem 3.1** (RGP upper bound of Subsampled Mechanisms). *Let $\mathcal{M} := \mathcal{A} \circ \text{Subsample}$ be a subsampled mechanism with a sampling rate $q \in (0, 1)$. If $\mathcal{A}$ satisfies $(k, \alpha, \tau_k^*(\alpha))$-bounded (resp. unbounded) RGP for $k \in \{0, 1, \dots, m\}$, then $\mathcal{M}$ satisfies $(m, \alpha, \tau_m(\alpha))$-bounded (resp. unbounded) RGP, where*

$$\tau_m(\alpha) = \frac{1}{\alpha - 1} \log \left( \sum_{k=0}^{m} \binom{m}{k} (1-q)^{m-k} q^k \exp\left( (\alpha - 1)\tau_k^*(\alpha) \right) \right).$$

The above theorem generally applies to all subsampled mechanisms with a constant sampling rate. To apply this upper bound to determine the exact RGP guarantee of a specific subsampled mechanism $\mathcal{M} = \mathcal{A} \circ \text{Subsample}$, a practitioner needs to determine the RGP guarantee $\tau_k^*(\alpha)$ of $\mathcal{A}$ for each group size $k \in \{0, 1, \dots, m\}$. In practice, the algorithm $\mathcal{A}$ is often derived from well-studied DP mechanisms, such as the Gaussian mechanism. Hence, we can usually bypass the existing RDP-to-RGP conversion method (Lemma 2.3), which is rather conservative as explained in Section 2.2, and directly derive a tight group privacy guarantee using properties of these well-studied mechanisms, elaborated in Section 4.

### 3.2 Proof of Our General RGP Bound

The proof of Theorem 3.1 is established by deriving upper bounds for both $D_\alpha(\mathcal{M}(D) \| \mathcal{M}(D'))$ and $D_\alpha(\mathcal{M}(D') \| \mathcal{M}(D))$, where $D$ and $D'$ are a pair of $m$-neighboring datasets. In what follows, we first establish the upper bound of the *unbounded* RGP guarantee for $\mathcal{M}$, and then derive the upper bound of the *bounded* RGP guarantee.

**Ensuring unbounded RGP.** Assume that $\mathcal{A}$ satisfies $(k, \alpha, \tau_k^*(\alpha))$-unbounded RGP for each $k \in \{0, 1, \dots, m\}$. Without loss of generality, consider $D$ and $D'$ as a pair of unbounded $m$-neighboring datasets such that $D \subset D'$, with $|D| = n$ and $|D'| - D| = m$. Let $\mathcal{B} := \{B \mid B \subseteq D\}$ denote the power set of $D$, and $\mathcal{J} := \{J \mid J \subseteq D' \setminus D\}$ represent the power set of $D' \setminus D$. Let $p_B$ denote the probability that $B$ is the outcome of the subsampling process $\text{Subsample}(D)$. Since the Subsample procedure places each record into the subset $B$ independently with a constant probability $q$, the values of $p_B$ are identical for both $\mathcal{M}(D)$ and $\mathcal{M}(D')$. Similarly, denote by $p_J$ the probability that $J$ is subsampled by $\mathcal{M}(D')$ from $D' \setminus D$. Then, the

output distributions of $\mathcal{M}(D)$ and $\mathcal{M}(D')$ can be expressed as

$$\mathcal{M}(D) \sim \sum_{B \in \mathcal{B}} p_B \mathcal{A}(B),$$

$$\mathcal{M}(D') \sim \sum_{B \in \mathcal{B}} p_B \sum_{J \in \mathcal{J}} p_J \mathcal{A}(B \cup J).$$

To establish the RGP guarantee for $\mathcal{M}$, we seek to upper bound $D_\alpha(\mathcal{M}(D) \| \mathcal{M}(D'))$ and $D_\alpha(\mathcal{M}(D') \| \mathcal{M}(D))$ simultaneously. Beginning with the term $D_\alpha(\mathcal{M}(D') \| \mathcal{M}(D))$, we have

$$D_\alpha(\mathcal{M}(D') \| \mathcal{M}(D)) = D_\alpha\left(\sum_{B \in \mathcal{B}} p_B \sum_{J \in \mathcal{J}} p_J \mathcal{A}(B \cup J) \middle\| \sum_{B \in \mathcal{B}} p_B \mathcal{A}(B)\right)$$

$$\leq \sup_B D_\alpha\left(\sum_{J \in \mathcal{J}} p_J \mathcal{A}(B \cup J) \middle\| \mathcal{A}(B)\right), \quad (3)$$

where the inequality follows from the joint quasi-convexity of Rényi divergence (see Corollary B.1).

Define $\widetilde{B} := \arg\max_{B \in \mathcal{B}} D_\alpha\left(\sum_{J \in \mathcal{J}} p_J \mathcal{A}(B \cup J) \middle\| \mathcal{A}(B)\right)$. Denote by $\boldsymbol{\mu}_J$ and $\boldsymbol{\mu}_0$ the probability density functions (pdfs) of $\mathcal{A}(\widetilde{B} \cup J)$ and $\mathcal{A}(\widetilde{B})$, respectively. Then by (3), we can further simplify the upper bound of $D_\alpha(\mathcal{M}(D') \| \mathcal{M}(D))$ as follows:

$$D_\alpha(\mathcal{M}(D') \| \mathcal{M}(D)) \leq D_\alpha\left(\sum_{J \in \mathcal{J}} p_J \boldsymbol{\mu}_J \middle\| \boldsymbol{\mu}_0\right). \quad (4)$$

The upper bound of the term $D_\alpha(\mathcal{M}(D) \| \mathcal{M}(D'))$ is derived similarly. Let $\widetilde{B'} := \arg\max_{B \in \mathcal{B}} D_\alpha\left(\mathcal{A}(B) \middle\| \sum_{J \in \mathcal{J}} p_J \mathcal{A}(B \cup J)\right)$ and denote by $\boldsymbol{\mu}'_J$ and $\boldsymbol{\mu}'_0$ the pdfs of $\mathcal{A}(\widetilde{B'} \cup J)$ and $\mathcal{A}(\widetilde{B'})$, respectively. It then holds that

$$D_\alpha(\mathcal{M}(D) \| \mathcal{M}(D')) \leq D_\alpha\left(\boldsymbol{\mu}'_0 \middle\| \sum_{J \in \mathcal{J}} p_J \boldsymbol{\mu}'_J\right). \quad (5)$$

Let $A_\alpha$ and $A'_\alpha$ denote the right hand side (RHS) of (4) and (5), respectively. The subsampled mechanism $\mathcal{M}$ then satisfies $(m, \alpha, \max\{A_\alpha, A'_\alpha\})$-unbounded RGP. We proceed by establishing an upper bound for $\max\{A_\alpha, A'_\alpha\}$. For the term $A_\alpha$, we have

$$A_\alpha = \frac{1}{\alpha-1} \mathbb{E}_{z \sim \boldsymbol{\mu}_0} \left[\left(\frac{\sum_{J \in \mathcal{J}} p_J \boldsymbol{\mu}_J(z)}{\boldsymbol{\mu}_0(z)}\right)^\alpha\right]$$

$$= \frac{1}{\alpha-1} \log \int_{\mathcal{Z}} \frac{\left(\sum_{J \in \mathcal{J}} p_J \boldsymbol{\mu}_J(z)\right)^\alpha}{\boldsymbol{\mu}_0(z)^{\alpha-1}} dz$$

$$\leq \frac{1}{\alpha-1} \log\left(\sum_{J \in \mathcal{J}} p_J \int_{\mathcal{Z}} \frac{\boldsymbol{\mu}_J(z)^\alpha}{\boldsymbol{\mu}_0(z)^{\alpha-1}} dz\right)$$

$$= \frac{1}{\alpha-1} \log\left(\sum_{J \in \mathcal{J}} p_J \exp\left((\alpha-1) D_\alpha(\boldsymbol{\mu}_J \| \boldsymbol{\mu}_0)\right)\right), \quad (6)$$

where the inequality follows from the convexity of the function $f(x) = x^\alpha$ for all $\alpha > 1$, and the last equality is derived from the definition of Rényi divergence.

Let $\mathcal{J}_k := \{J \in \mathcal{J} \mid |J| = k\}$. Because $\mathcal{A}$ satisfies $(k, \alpha, \tau_k^*(\alpha))$-unbounded RGP, it is established that $D_\alpha(\boldsymbol{\mu}_J \| \boldsymbol{\mu}_0) \leq \tau_k^*(\alpha)$ for all

$J \in \mathcal{J}_k$. Consequently, it follows that

$$(6) = \frac{1}{\alpha-1} \log\left(\sum_{k=0}^{m} \sum_{J \in \mathcal{J}_k} p_J \exp\left((\alpha-1) D_\alpha(\boldsymbol{\mu}_J \| \boldsymbol{\mu}_0)\right)\right)$$

$$\leq \frac{1}{\alpha-1} \log\left(\sum_{k=0}^{m} p_k \exp\left((\alpha-1) \tau_k^*(\alpha)\right)\right), \quad (7)$$

where $p_k := \sum_{J \in \mathcal{J}_k} p_J$.

We now proceed to upper bound the term $A'_\alpha$. Because Rényi divergence is convex in its second term (see Lemma B.2), we immediately obtain

$$A'_\alpha = D_\alpha\left(\boldsymbol{\mu}'_0 \middle\| \sum_J p_J \boldsymbol{\mu}'_J\right) \leq \sum_J p_J D_\alpha\left(\boldsymbol{\mu}'_0 \middle\| \boldsymbol{\mu}'_J\right) \leq \sum_{k=0}^{m} p_k \tau_k^*(\alpha). \quad (8)$$

Denote by $\overline{A_\alpha}$ and $\overline{A'_\alpha}$ the RHS of (7) and (8), respectively. Given the concavity of the logarithm function, we have

$$\overline{A_\alpha} = \frac{1}{\alpha-1} \log\left(\sum_{k=0}^{m} p_k \exp\left((\alpha-1)\tau_k^*(\alpha)\right)\right)$$

$$\geq \frac{1}{\alpha-1} \sum_{k=0}^{m} p_k \log\left(\exp\left((\alpha-1)\tau_k^*(\alpha)\right)\right) = \sum_{k=0}^{m} p_k \tau_k^*(\alpha) = \overline{A'_\alpha}.$$

Therefore, $\max\{A_\alpha, A'_\alpha\}$ can be upper bounded as:

$$\max\{A_\alpha, A'_\alpha\} \leq \max\{\overline{A_\alpha}, \overline{A'_\alpha}\}$$

$$= \overline{A_\alpha} = \frac{1}{\alpha-1} \log\left(\sum_{k=0}^{m} p_k \exp\left((\alpha-1)\tau_k^*(\alpha)\right)\right). \quad (9)$$

Recall that in the underlying subsampled mechanism, each record in $D' \setminus D$ is randomly selected into the input subset independently with a fixed probability $q$. Accordingly, the number of records in $J$ follows a binomial distribution with $m$ trials and success probability $q$. Hence, we have

$$p_k = \binom{m}{k}(1-q)^{m-k} q^k. \quad (10)$$

Substituting (10) into (9) yields the privacy guarantee with respect to unbounded RGP.

**Ensuring *bounded* RGP.** We slightly abuse the notations and let $D$ and $D'$ be a pair of $m$-bounded neighboring datasets with $|D| = |D'| = n$. Without loss of generality, we assume that $D$ and $D'$ differ in the last $m$ records. Suppose $\mathcal{A}$ satisfies $(k, \alpha, \tau_k^*(\alpha))$-bounded RGP for each $k \in \{0, 1, \ldots, m\}$. Define $\mathcal{I} := \{I \mid I \subseteq \{1, 2, \ldots, n\}\}$ as the power set of the indices $\{1, 2, \ldots, n\}$ and let $D_I := \{\boldsymbol{x}_i \mid i \in I, \boldsymbol{x}_i \in D\}$ be the set of records in $D$ indexed by $I$. Denote by $p_I$ the probability that the subset $D_I$ is the outcome of Subsample($D$). Consequently, the output distributions of $\mathcal{M}(D)$ and $\mathcal{M}(D')$ can be expressed as:

$$\mathcal{M}(D) \sim \sum_{I \in \mathcal{I}} p_I \mathcal{A}(D_I) \quad \text{and} \quad \mathcal{M}(D') \sim \sum_{I \in \mathcal{I}} p_I \mathcal{A}(D'_I).$$

Let $\nu_I$ and $\nu'_I$ denote the pdfs of $\mathcal{A}(D_I)$ and $\mathcal{A}(D'_I)$. Then

$$
\begin{aligned}
D_\alpha(\mathcal{M}(D)\|\mathcal{M}(D')) &= D_\alpha\left(\sum_{I\in\mathcal{I}} p_I \nu_I \,\middle\|\, \sum_{I\in\mathcal{I}} p_I \nu'_I\right) \\
&= \frac{1}{\alpha-1}\log \int_{\mathcal{Z}} \frac{\left(\sum_{I\in\mathcal{I}} p_I \nu_I(z)\right)^\alpha}{\left(\sum_{I\in\mathcal{I}} p_I \nu'_I(z)\right)^{\alpha-1}}\, \mathrm{d}z \\
&\leq \frac{1}{\alpha-1}\log\left(\sum_{I\in\mathcal{I}} p_I \int_{\mathcal{Z}} \frac{\nu_I(z)^\alpha}{\nu'_I(z)^{\alpha-1}}\, \mathrm{d}z\right) \\
&= \frac{1}{\alpha-1}\log\left(\sum_{I\in\mathcal{I}} p_I \exp\big((\alpha-1)D_\alpha(\nu_I\|\nu'_I)\big)\right),
\end{aligned}
$$

where the inequality is due to a joint convexity argument (see Lemma B.4). Similarly, we have

$$
D_\alpha(\mathcal{M}(D')\|\mathcal{M}(D)) \leq \frac{1}{\alpha-1}\log\left(\sum_{I\in\mathcal{I}} p_I \exp\big((\alpha-1)D_\alpha(\nu'_I\|\nu_I)\big)\right).
$$

Let $\mathcal{I}_k := \{I\in\mathcal{I} \mid |D_I\setminus D'_I| = k\}$ denote the set of $I$ such that $D_I$ and $D'_I$ differ by $k$ records, then we have $\sum_{I\in\mathcal{I}_k} p_I = p_k$. Since $\mathcal{A}$ satisfies $(k,\alpha,\tau^*_k(\alpha))$-bounded RGP, both $D_\alpha(\nu'_I\|\nu_I)$ and $D_\alpha(\nu_I\|\nu'_I)$ are upper bounded by $\tau^*_k(\alpha)$ for all $I\in\mathcal{I}_k$. Therefore, it follows that

$$
\begin{aligned}
&(\alpha-1)\cdot\max\left\{D_\alpha(\mathcal{M}(D)\|\mathcal{M}(D')), D_\alpha(\mathcal{M}(D')\|\mathcal{M}(D))\right\} \\
&\leq \log\left(\sum_{k=0}^m \sum_{I\in\mathcal{I}_k} p_I \exp\big((\alpha-1)\cdot\max\left\{D_\alpha(\nu_I\|\nu'_I), D_\alpha(\nu'_I\|\nu_I)\right\}\big)\right) \\
&\leq \log\left(\sum_{k=0}^m p_k \exp\big((\alpha-1)\tau^*_k(\alpha)\big)\right),
\end{aligned}
$$

thereby completing the proof for bounded group privacy after substituting the $p_k$ into (10).

Note that in the above analysis, we implicitly assume that the probability density functions $\mu_k$ and $\nu_I$ are continuous for simplicity. In discrete scenarios, one can readily confirm the validity of the above result by substituting integrals with summations. Therefore, Theorem 3.1 is applicable to both continuous and discrete scenarios.

### 3.3 Tightness of Our General RGP Bound

Next, we demonstrate that the RGP bound in Theorem 3.1 is asymptotic optimal. Specifically, we first establish a lower bound for the privacy cost of subsampled mechanisms under RGP, and then show that the established lower bound and the upper bound in Theorem 3.1 match up to an additive constant factor. This lower bound is obtained by constructing a pair of binary and one-dimensional $m$-neighboring datasets $D$ and $D'$, as follows:

$$
D = \{0, 0, \ldots, 0\} \quad \text{and} \quad D' = D \cup \underbrace{\{1, 1, \ldots, 1\}}_{m\text{ records}}. \tag{11}
$$

Let $\mathcal{A}$ be a Gaussian mechanism that takes a binary dataset as the input and output the sum of its records in a differentially private manner, i.e., $\mathcal{A}(D) = \sum_{x\in D} x + \mathcal{N}(0,\sigma^2)$, where $\mathcal{N}(0,\sigma^2)$ denotes the Gaussian noise with mean 0 and variance $\sigma^2$. Define $\mathcal{M}(\cdot) := \mathcal{A}\circ \mathrm{Subsample}(\cdot)$ as a subsampled Gaussian mechanism with a sampling rate $q$, and denote by $\mu_k$ the pdf of the distribution $\mathcal{N}(k,\sigma^2)$ for $k\in\mathbb{N}$, then we have

$$
\mathcal{M}(D)\sim\mu_0 \quad \text{and} \quad \mathcal{M}(D')\sim\sum_{k=0}^m p_k\mu_k,
$$

where $p_k = \binom{m}{k}(1-q)^{m-k}q^k$.

Now suppose that $\mathcal{A}$ satisfies $(k,\alpha,\tau^*_k(\alpha))$-unbounded RGP and $\mathcal{M}$ satisfies $(m,\alpha,\tau_m(\alpha))$-unbounded RGP. This immediately implies that $\tau_m(\alpha)\geq D_\alpha(\mathcal{M}(D')\|\mathcal{M}(D))$. Hence, we obtain:

$$
\begin{aligned}
\tau_m(\alpha) &\geq \frac{1}{\alpha-1}\log\mathbb{E}_{\mu_0}\left[\left(\frac{\sum_{k=0}^m p_k\mu_k}{\mu_0}\right)^\alpha\right] \\
&\geq \frac{1}{\alpha-1}\log\mathbb{E}_{\mu_0}\left[\sum_{k=0}^m p_k^\alpha\left(\frac{\mu_k}{\mu_0}\right)^\alpha\right] \\
&= \frac{1}{\alpha-1}\log\left(\sum_{k=0}^m p_k^\alpha\mathbb{E}_{\mu_0}\left[\left(\frac{\mu_k}{\mu_0}\right)^\alpha\right]\right) \\
&= \frac{1}{\alpha-1}\log\left(\sum_{k=0}^m p_k^\alpha\exp\big((\alpha-1)\tau^*_k(\alpha)\big)\right), \tag{12}
\end{aligned}
$$

where the last equality holds because a direct calculation verifies that $\mathbb{E}_{\mu_0}\left[(\mu_k/\mu_0)^\alpha\right] = \alpha k^2/2\sigma^2$, which attains the RGP guarantee of the Gaussian mechanism $\mathcal{A}$ (see Lemma 4.2). Define

$$
\begin{aligned}
\Xi^+_{\alpha,m,q} &:= \sum_{k=0}^m \binom{m}{k}(1-q)^{m-k}q^k \exp\big((\alpha-1)\tau^*_k(\alpha)\big), \\
\Xi^-_{\alpha,m,q} &:= \sum_{k=0}^m \left(\binom{m}{k}(1-q)^{m-k}q^k\right)^\alpha \exp\big((\alpha-1)\tau^*_k(\alpha)\big),
\end{aligned}
$$

then by Theorem 3.1 and (12), the upper and lower bounds of the group privacy guarantee of $\mathcal{M}$ can be expressed as $\frac{1}{\alpha-1}\log\Xi^+_{\alpha,m,q}$ and $\frac{1}{\alpha-1}\log\Xi^-_{\alpha,m,q}$, respectively.

We proceed to compare the terms $\Xi^+_{\alpha,m,q}$ and $\Xi^-_{\alpha,m,q}$. Note that

$$
\sum_{k=0}^m \left(\binom{m}{k}(1-q)^{m-k}q^k\right)^\alpha \geq \left(\binom{m}{0}(1-q)^m q^0\right)^\alpha
$$
$$
= (1-q)^{m\alpha} \geq 1 - qm\alpha, \tag{13}
$$

where the last inequality follows from the Bernoulli's inequality. Define $c_\alpha := \max\{(\alpha-1)\tau^*_k(\alpha)\}_{k=0}^m$. With an appropriate setting for the injected noise to satisfy group privacy (e.g., $\sigma = \Theta(m)$ in Gaussian mechanisms, elaborate later in Section 4.1), $c_\alpha$ becomes a constant depending solely on $\alpha$. Thus, we derive:

$$
\begin{aligned}
\Xi^+_{\alpha,m,q} - \Xi^-_{\alpha,m,q} &= \sum_{k=0}^m \left(p_k - p_k^\alpha\right) e^{((\alpha-1)\tau^*_k(\alpha))} \\
&\leq e^{c_\alpha}\left(\sum_{k=0}^m p_k - \sum_{k=0}^m p_k^\alpha\right) = e^{c_\alpha}\left(1 - \sum_{k=0}^m p_k^\alpha\right) \leq e^{c_\alpha} qm\alpha,
\end{aligned}
$$

where the last inequality follows from (13). By applying the mean value theorem to the logarithm function, we can verify that the upper and lower bounds, i.e., $\frac{1}{\alpha-1}\log\Xi^+_{\alpha,m,q}$ and $\frac{1}{\alpha-1}\log\Xi^-_{\alpha,m,q}$, match up to an additive factor $O(e^{c_\alpha}qm\alpha)$. Setting $qm = O(1)$ and treating terms influenced by $\alpha$ as constants implies that the upper and lower bounds match up to an additive constant factor, implying the asymptotic tightness of our bound.

**Remark.** As we show soon in Section 4.1, the RGP upper and lower bounds discussed above for the DP-SGD algorithm match up to an additive factor of $O(\alpha e^{\alpha^2})$, provided that the noise scale $\sigma = \Theta(m)$ and $qm = O(1)$. Further, although the above description uses the subsampled Gaussian mechanism as an example, the analysis of the lower bound can be adapted for other mechanisms, which is detailed in Appendix D.

While asymptotic results provide valuable insights on the optimality of our general RGP bound, for practical applications, we also need concrete bounds for calibrating the noise levels. In the next section, we further derive closed-form group privacy bounds for a range of widely-used subsampled mechanisms.

## 4 APPLICATIONS

This section presents closed-form group privacy bounds for various privacy-preserving subsampled mechanisms. In what follows, we denote by $f$ the algorithm intended for privatization under RGP.

We first present a lemma which is instrumental in deriving tight RGP bounds for non-subsampled mechanisms, i.e., the term $\tau_k^*(\alpha)$ in Theorem 3.1. This lemma serves as a useful tool for deriving closed-form RGP guarantees for subsampled Gaussian, Laplace, and Skellam mechanisms, detailed in the following subsections.

**Lemma 4.1.** *Let $f : \mathcal{D} \mapsto \mathbb{R}^d$ be a function that maps a dataset to a $d$-dimensional vector and let $\| \cdot \|$ be an arbitrary norm. If $\|f(D) - f(\hat{D})\| \leq C$ holds for any pair of bounded (resp. unbounded) 1-neighboring datasets $D$ and $\hat{D}$ that differ by one record, then for all $k \in \mathbb{N}$, it holds that $\|f(D) - f(D')\| \leq kC$ for any pair of bounded (resp. unbounded) $k$-neighboring datasets $D$ and $D'$.*

PROOF. The lemma is established by applying the triangle inequality of norms and the principle of mathematical induction. □

### 4.1 Subsampled Gaussian Mechanism

We now establish the closed-form RGP bound for the subsampled Gaussian mechanism, which is a fundamental component in many popular privacy-preserving applications, most notably DP-SGD [1]. Let $f : \mathcal{D} \mapsto \mathbb{R}^d$ denote the algorithm for which $\|f(D) - f(\hat{D})\|_2 \leq C$ holds for all pairs of datasets differing by one record. The Gaussian mechanism that privatizes the algorithm $f$ is defined as follows:

$$\mathcal{A}(D) = f(D) + \mathcal{N}(\mathbf{0}, C^2 \sigma^2 \mathbb{I}^d),$$

where $\mathcal{N}(\mathbf{0}, C^2 \sigma^2 \mathbb{I}^d)$ represents the spherical $d$-dimensional Gaussian noise with per-coordinate variance $C^2 \sigma^2$.

For any pair of $k$-neighboring dataset $D$ and $D'$, the Rényi divergence between $\mathcal{A}(D)$ and $\mathcal{A}(D')$ can be upper bounded as:

$$
\begin{aligned}
&D_\alpha \Big( \mathcal{N}\big( f(D), C^2 \sigma^2 \mathbb{I}^d \big) \Big\| \mathcal{N}\big( f(D'), C^2 \sigma^2 \mathbb{I}^d \big) \Big) \\
&\overset{(a)}{=} D_\alpha \Big( \mathcal{N}\big( f(D) - f(D'), C^2 \sigma^2 \mathbb{I}^d \big) \Big\| \mathcal{N}\big( \mathbf{0}, C^2 \sigma^2 \mathbb{I}^d \big) \Big) \\
&\overset{(b)}{\leq} \sup_{\|\mathbf{v}\|_2 \leq kC} D_\alpha \Big( \mathcal{N}\big( \mathbf{v}, C^2 \sigma^2 \mathbb{I}^d \big) \Big\| \mathcal{N}\big( \mathbf{0}, C^2 \sigma^2 \mathbb{I}^d \big) \Big) \\
&\overset{(c)}{=} \sup_{\|\mathbf{v}\|_2 \leq kC} \sum_{i=1}^d D_\alpha \Big( \mathcal{N}\big( \mathbf{v}[i], C^2 \sigma^2 \big) \Big\| \mathcal{N}\big( \mathbf{0}, C^2 \sigma^2 \big) \Big) \\
&\overset{(d)}{=} \sup_{\|\mathbf{v}\|_2 \leq kC} \sum_{i=1}^d \frac{\alpha \mathbf{v}[i]^2}{2 C^2 \sigma^2} = \sup_{\|\mathbf{v}\|_2 \leq kC} \frac{\alpha \|\mathbf{v}\|_2^2}{2 C^2 \sigma^2} = \frac{\alpha k^2}{2 \sigma^2}, \quad (14)
\end{aligned}
$$

where (a) follows from the invariance of Rényi divergence under invertible transformations, which is a variation of the more general data processing inequality [52]; (b) is derived from Lemma 4.1; (c) follows from the additivity of Rényi divergence (see Lemma B.3); (d) follows from the closed-form Rényi divergence between Gaussian distributions (see Lemma B.5). Here, $\mathbf{v}[i]$ denotes the $i$-th element of vector $\mathbf{v}$. It is worth noting that the upper bound in (14), i.e., $\alpha k^2 / 2\sigma^2$, can be attained by specific $f$ and $D$, affirming the tightness of our analysis.

Accordingly, we establish the RGP guarantee for the Gaussian mechanism as follows:

**Lemma 4.2.** *Let $\mathcal{A}$ be a Gaussian mechanism defined as above, it holds that $\mathcal{A}$ satisfies $(k, \alpha, \tau = \frac{\alpha k^2}{2\sigma^2})$-RGP for all $k \in \mathbb{N}$.*

Comparing the above result with the generic RDP-to-RGP conversion in Lemma 2.3, observe that the former applies to any group size, whereas the latter is limited to the case where the group size is a power of two, i.e., $2^c$ for $c \in \mathbb{N}$. Further, Lemma 4.2 above leads to the same value of $\alpha$ in both the RDP (i.e., by setting $k = 1$) and RGP ($k > 1$), whereas in Lemma 2.3, the converted RGP uses a lower $\alpha' = \frac{\alpha}{2^c}$. If, in Lemma 4.2, we also aim to satisfy RGP with this smaller $\alpha'$, with $k = 2^c$, we would have $\tau = \frac{\alpha' k^2}{\sigma^2} = \frac{\alpha k}{2\sigma^2}$, which grows linearly with the group size $k$, rather than with $k^{1.58}$ as discussed in Section 2.2. Hence, the above lemma provides a more refined privacy analysis compared to the naive RDP-to-RGP conversion approach.

Next, we extend our refined, closed-form privacy cost analysis to the subsampled Gaussian mechanism. Combining Lemma 4.2 with Theorem 3.1, we arrive at the following theorem.

**Theorem 4.1.** *Let $\mathcal{M} := \mathcal{A} \circ \text{Subsample}$ denote the subsampled Gaussian mechanism with sampling rate $q \in (0, 1)$, where $\mathcal{A}$ is the Gaussian mechanism defined above. Then, $\mathcal{M}$ satisfies $(m, \alpha, \tau_m(\alpha))$-RGP, with*

$$\tau_m(\alpha) = \frac{1}{\alpha - 1} \log \left( \sum_{k=0}^m \binom{m}{k} (1-q)^{m-k} q^k \exp \left( \frac{(\alpha - 1)\alpha k^2}{2\sigma^2} \right) \right).$$

The following theorem shows that with respect to the subsampled Gaussian mechanism, the above result achieves an asymptotically improved privacy guarantee compared to baseline solution of RDP-to-RGP conversion through Lemma 2.3.

**Theorem 4.2.** *Consider a subsampled Gaussian mechanism $\mathcal{M}$ with a sampling rate $q \in (0, 1)$ and variance $\sigma^2$. Let $\alpha > 1$ be any* *integer. We denote by $(m, \alpha, \tau'_m(\alpha))$ and $(m, \alpha, \tau_m(\alpha))$ the unbounded RGP guarantees of $\mathcal{M}$ derived using Lemma 2.3 and Theorem 4.1, respectively. If $\sigma = \Theta(m)$ and $qm > 1$, our bound $\tau_m(\alpha)$ saves a multiplicative factor of $\Theta(m^{\log_2 1.5})$ compared to $\tau'_m(\alpha)$.*

PROOF SKETCH. The complete proof is deferred to Appendix C.1. Here we present the proof sketch as follows.

Both the RGP bounds derived by the baseline solution and our solution can be expressed in a binomial expression-like form that includes exponential terms. These terms are rather complicated to compare directly. To address this challenge, we approximate these exponential terms using polynomials via Taylor's theorem. This approximation allows the RGP bounds to be asymptotically expressed as the summation of high moments of binomial distributions, yielding simpler and clearer bounds. Accordingly, we derive asymptotic improvements based on these refined bounds. □

**Remark.** The above theorem indicates that our RGP bound significantly enhances the privacy guarantee for the DP-SGD algorithm, on the condition that $\sigma = \Theta(m)$ and $qm > 1$. Note that these two conditions align with the conditions in the remark toward the end of Section 3.3, signifying that our bound is provably tight for DP-SGD while saving a $\Theta(m^{\log_2 1.5}) \approx \Theta(m^{0.58})$ multiplicative factor[4] in terms of RGP guarantee compared to the baseline method.

### 4.2 Subsampled Laplace Mechanism

The Laplace mechanism [20] is among the most widely used mechanisms for achieving DP in numerous practical applications [9, 18, 21, 60]. Let $f : \mathcal{D} \mapsto \mathbb{R}^d$ denote the algorithm for which $\|f(D) - f(\hat{D})\|_1 \leq C$ is satisfied for all pairs of neighboring datasets that differ by one record. The Laplace mechanism is defined as

$$\mathcal{A}(D) = f(D) + \mathsf{Lap}(\mathbf{0}, Cb\mathbb{I}^d),$$

where $\mathsf{Lap}(\mathbf{0}, Cb\mathbb{I}^d)$ denotes the $d$-dimensional Laplace noise with per-coordinate scale factor $Cb$, i.e., each coordinate is independently drawn from the one-dimensional Laplace distribution $\mathsf{Lap}(0, Cb)$.

We can upper bound the Rényi divergence between $\mathcal{A}(D)$ and $\mathcal{A}(D')$ for any pair of $k$-neighboring datasets $D$ and $D'$ as follows:

$$
\begin{aligned}
&D_\alpha(\mathsf{Lap}(f(D), Cb\mathbb{I}^d) \| \mathsf{Lap}(f(D'), Cb\mathbb{I}^d))\\
&\stackrel{(a)}{=} D_\alpha\left(\mathsf{Lap}(f(D) - f(D'), Cb\mathbb{I}^d) \Big\| \mathsf{Lap}(\mathbf{0}, Cb\mathbb{I}^d)\right)\\
&\stackrel{(b)}{\leq} \sup_{\|\mathbf{v}\|_1 \leq kC} \sum_{i=1}^d D_\alpha(\mathsf{Lap}(\mathbf{v}[i], Cb) \| \mathsf{Lap}(0, Cb))\\
&\stackrel{(c)}{=} \sup_{\|\mathbf{v}\|_1 \leq kC} \frac{1}{\alpha - 1} \sum_{i=1}^d \log\left\{\frac{\alpha}{2\alpha - 1} \exp\left(\frac{(\alpha - 1)\mathbf{v}[i]}{Cb}\right)\right.\\
&\qquad\qquad\qquad\left. + \frac{\alpha - 1}{2\alpha - 1} \exp\left(\frac{-\alpha \mathbf{v}[i]}{Cb}\right)\right\}, \quad (15)
\end{aligned}
$$

---

[4] Based on our analysis in Appendix C.1, we can derive that when the Gaussian variance satisfies $\sigma = \Theta(m)$, our RGP guarantee is upper bounded by $\alpha q^2/2$ as $m$ becomes large, while the baseline RGP guarantee is lower bounded by $m^{\log_2 1.5} \alpha q^2/4 \approx m^{0.58} \alpha q^2/4$. Therefore, a more accurate bound on the improvement is $m^{0.58}/2$. In Theorem 4.2, we use the big-oh notation to demonstrate the asymptotic improvement of our result, simplifying the expression by ignoring constant factors.

where (a) follows from the invariance of Rényi divergence under invertible transformations; (b) follows from Lemma 4.1 and the additivity of Rényi divergence (see Lemma B.3); (c) follows from the closed-form Rényi divergence between Laplace distributions (see Lemma B.6).

The subsequent task is to derive a closed-form upper bound of (15). Unlike the Gaussian mechanism, the expression in (15) is rather complicated, and thus identifying its maximum is not straightforward. To address this, we formulate the following constrained optimization problem, which is equivalent to determining the maximum of (15):

$$
\begin{aligned}
\text{maximize}_{\{x_i\}} \quad &\sum_{i=1}^d \log\left\{\frac{\alpha}{2\alpha - 1} \exp\left(\frac{(\alpha - 1)x_i}{Cb}\right)\right.\\
&\qquad\qquad\left. + \frac{\alpha - 1}{2\alpha - 1} \exp\left(\frac{-\alpha x_i}{Cb}\right)\right\}\\
\text{subject to} \quad &\sum_{i=1}^d |x_i| \leq kC. \quad (16)
\end{aligned}
$$

Next, we introduce the following lemmas, which are crucial tools for solving the above constraint optimization problem.

**Lemma 4.3.** *The function $f(x) = \log\left(c_1 e^{\beta_1 x} + c_2 e^{-\beta_2 x}\right)$ is convex for $c_1, c_2, \beta_1, \beta_2 \in (0, \infty)$.*

PROOF. See Appendix C.2. □

**Lemma 4.4** (Bauer's Maximum Principle [10]). *A maximum of a convex function over a closed and bounded convex set is achieved at an extreme point.*

We are now ready to solve the optimization problem (15). Note that for all $\alpha > 1$, we have $\frac{\alpha}{2\alpha - 1} > 0$, $\frac{\alpha - 1}{2\alpha - 1} > 0$, $\frac{\alpha - 1}{Cb} > 0$, and $\frac{\alpha}{Cb} > 0$. Thus, by Lemma 4.3, the objective function of (16) is convex. Moreover, the domain of this objective function is a convex polyhedron (an $L_1$-ball), with vertices constituting its extreme points. Consequently, by Lemma 4.4, the maximum of (15) is attained at $\mathbf{v} = kC \cdot \mathbf{e}$ for some vector $\mathbf{e}$ in the standard basis. Denote by $\mathbf{e}_i$ the $i$-th element of the vector $\mathbf{e}$, then we have

$$
\begin{aligned}
(15) &\leq \frac{1}{\alpha - 1} \sum_{i=1}^d \log\left\{\frac{\alpha}{2\alpha - 1} \exp\left(\frac{(\alpha - 1)kC\mathbf{e}_i}{Cb}\right)\right.\\
&\qquad\qquad\left. + \frac{\alpha - 1}{2\alpha - 1} \exp\left(\frac{-\alpha kC\mathbf{e}_i}{Cb}\right)\right\}\\
&= \frac{1}{\alpha - 1} \log\left\{\frac{\alpha}{2\alpha - 1} \exp\left(\frac{(\alpha - 1)k}{b}\right) + \frac{\alpha - 1}{2\alpha - 1} \exp\left(\frac{-\alpha k}{b}\right)\right\}. \quad (17)
\end{aligned}
$$

Substituting (17) into Theorem 3.1 establishes the following group privacy guarantee for the subsampled Laplace mechanism:

**Theorem 4.3.** *Let $\mathcal{M} := \mathcal{A} \circ \text{Subsample}$ be a subsampled Laplace mechanism with sampling rate $q \in (0, 1)$, where $\mathcal{A}$ is the Laplace mechanism defined above. Then, $\mathcal{M}$ satisfies $(m, \alpha, \tau_m(\alpha))$-RGP, where*

$$\tau_m(\alpha) = \frac{1}{\alpha - 1} \log\left(\sum_{k=0}^m \binom{m}{k}(1-q)^{m-k} q^k \phi_k^{\mathsf{Lap}}(\alpha)\right),$$

with

$$\Phi_k^{\mathsf{Lap}}(\alpha) = \frac{\alpha}{2\alpha - 1} \exp\left(\frac{(\alpha - 1)k}{b}\right) + \frac{\alpha - 1}{2\alpha - 1} \exp\left(\frac{-\alpha k}{b}\right).$$

**Remark.** The above result is not only interesting for RGP, but also for the RDP setting. Before this work, the best known RDP guarantee to our knowledge for $d$-dimensional Laplace mechanism (i.e., Theorem 11 in [14]) accumulates privacy cost across all $d$-dimensions, i.e., by applying privacy composition $d$ times, which incurs a multiplicative factor of $d$. In contrast, our analysis reduces the problem to a simpler, one-dimensional case by carefully examining the conditions leading to the worst-case scenario. This approach allows us to derive a privacy guarantee without using composition, thus achieving a significant improvement.

Meanwhile, it is also important to note that the fact that our analysis bypasses privacy composition does not imply a violation of the well-established $\Omega(\sqrt{d})$ error bound of $\mathcal{A}$ from [17]. The Laplace noise is proportional to the $L_1$ norm of the output, and thus, the error associated with the Laplace mechanism still (implicitly) depends on the output dimension $d$.

### 4.3 Subsampled Skellam Mechanism

The Skellam mechanism [3, 8] applies discrete noise following a symmetric Skellam distribution to achieve DP. This mechanism is particularly useful in federated learning frameworks operating with multi-party computation (MPC) protocols [11, 56, 61], since these protocols use modular arithmetic as the fundamental cryptographic primitive [59], hence limiting their computation in finite fields. As a result, common mechanisms that typically inject real-valued noise, including the Gaussian and the Laplace mechanisms discussed in previous subsections, are inherently incompatible with these protocols. We refer interested readers to [3, 8] for more details.

Let $\mathsf{Sk}(z, \mu)$ denote the one-dimensional symmetric Skellam distribution with mean $z \in \mathbb{Z}$ and variance $2\mu$. Denote by $f : \mathcal{D} \mapsto \mathbb{Z}^d$ the algorithm that maps an input dataset to an integer-valued vector such that $\|f(D) - f(\hat{D})\|_1 \leq C$ for pair of datasets $D$ and $\hat{D}$ that differ by one record. The Skellam mechanism is defined as follows:

$$\mathcal{A}(D) = f(D) + \mathsf{Sk}(\mathbf{0}, C^2\mu\mathbb{I}^d),$$

where $\mathsf{Sk}(\mathbf{0}, C^2\mu\mathbb{I}^d)$ denotes the multi-dimensional Skellam distribution with each coordinate distributed independently as $\mathsf{Sk}(0, C^2\mu)$.

The Rényi divergence $D_\alpha(\mathcal{A}(D)\|\mathcal{A}(D'))$ for any pair of $k$-neighboring datasets $D$ and $D'$ can be bounded as:

$$D_\alpha(\mathsf{Sk}(f(D), C^2\mu\mathbb{I}^d)\|\mathsf{Sk}(f(D'), C^2\mu\mathbb{I}^d))$$

$$\overset{(a)}{=} D_\alpha(\mathsf{Sk}(f(D) - f(D'), C^2\mu\mathbb{I}^d)\|\mathsf{Sk}(\mathbf{0}, C^2\mu\mathbb{I}^d))$$

$$\overset{(b)}{\leq} \sup_{\|\mathbf{v}\|_1 \leq kC} \sum_{i=1}^{d}\left(\frac{\alpha\mathbf{v}[i]^2}{2C^2\mu} + \min\left\{\frac{(2\alpha - 1)\mathbf{v}[i]^2 + 6|\mathbf{v}[i]|}{4C^4\mu^2}, \frac{3|\mathbf{v}[i]|}{2C^2\mu}\right\}\right)$$

$$\overset{(c)}{\leq} \sup_{\|\mathbf{v}\|_1 \leq kC}\left(\frac{\alpha\|\mathbf{v}\|_2^2}{2C^2\mu} + \min\left\{\frac{(2\alpha - 1)\|\mathbf{v}\|_2^2 + 6\|\mathbf{v}\|_1}{4C^4\mu^2}, \frac{3\|\mathbf{v}\|_1}{2C^2\mu}\right\}\right)$$

$$\overset{(d)}{\leq} \frac{\alpha k^2}{2\mu} + \min\left\{\frac{(2\alpha - 1)k^2C + 6k}{4C^3\mu^2}, \frac{3k}{2C\mu}\right\}, \quad (18)$$

where (a) follows from the invariance of Rényi divergence under invertible transformations; (b) is derived using Lemma 4.1, the closed-form Rényi divergence between symmetric Skellam distributions (see Lemma B.7), and the additivity of Rényi divergence (see Lemma B.3); (c) follows from the definitions of $L_1$ and $L_2$ norms; (d) follows from the fact that $\|\mathbf{v}\|_1 \leq C$ implies $\|\mathbf{v}\|_2 \leq C$. Combining (18) with Theorem 3.1 leads to the following group privacy guarantee for subsampled Skellam mechanisms:

**Theorem 4.4.** *Let $\mathcal{M}$ be the subsampled Skellam mechanism as defined above. $\mathcal{M}$ then satisfies $(m, \alpha, \tau_m(\alpha))$-RGP, where*

$$\tau_m(\alpha) = \frac{1}{\alpha - 1}\log\left(\sum_{k=0}^{m}\binom{m}{k}(1-q)^{m-k}q^k \exp\left((\alpha - 1)\Phi_k^{\mathsf{Sk}}(\alpha)\right)\right),$$

*with*

$$\Phi_k^{\mathsf{Sk}}(\alpha) = \frac{\alpha k^2}{2\mu} + \min\left\{\frac{(2\alpha - 1)k^2C + 6k}{4C^3\mu^2}, \frac{3k}{2C\mu}\right\}.$$

### 4.4 Subsampled Randomized Response

The randomized response (RR) mechanism [54] allow for the collection of statistical information about sensitive datasets while ensuring a DP guarantee. Given a predicate function $f : \mathcal{D} \mapsto \{0, 1\}$, the randomized response mechanism $\mathcal{A}$ that privatizes $f$ with privacy parameter $p \in (0.5, 1)$ is defined as follows:

$$\mathcal{A}(D) = \begin{cases} f(D), & \text{w.p.} \quad p \\ 1 - f(D), & \text{w.p.} \quad 1 - p. \end{cases} \quad (19)$$

Note that the RR mechanism provides a local DP guarantee [22], ensuring that the outputs of RR mechanisms are indistinguishable for any pair of datasets that differ by any number of records. Therefore, the RR mechanism satisfies $(m, \alpha, \tau_1^*(\alpha))$-RGP for any $m \in \mathbb{N}$. According to Lemma B.8, it holds that

$$\tau_1^*(\alpha) = \frac{1}{\alpha - 1}\log\left(\frac{p^\alpha}{(1-p)^{\alpha-1}} + \frac{(1-p)^\alpha}{p^{\alpha-1}}\right). \quad (20)$$

In addition, a direct calculation yields $\tau_0^*(\alpha) = 0$. Consequently, we can establish the following group privacy guarantee for the subsampled RR mechanism:

**Theorem 4.5.** *The subsampled RR mechanism with sampling rate $q$ and privacy parameter $p$ satisfies $(m, \alpha, \tau_m(\alpha))$-group RDP, where*

$$\tau_m(\alpha) = \frac{1}{\alpha - 1}\log\left((1-q)^m + \left(1 - (1-q)^m\right) \cdot \Phi^{\mathsf{RR}}(\alpha)\right),$$

*where*

$$\Phi^{\mathsf{RR}}(\alpha) = \frac{p^\alpha}{(1-p)^{\alpha-1}} + \frac{(1-p)^\alpha}{p^{\alpha-1}}$$

The following lemma shows that our group privacy bound for the RR mechanism achieves a provable improvement over naive bound $\tau_1^*(\alpha)$ defined in Eq. (20).

**Lemma 4.5.** *For the RR mechanism, it holds that $\lim_{m\to\infty}\tau_m(\alpha) = \tau_1^*(\alpha)$. In addition, we have $\tau_m(\alpha) < \tau_1^*(\alpha)$ for all $m \in \mathbb{N}$.*

Proof. See Appendix C.3. □

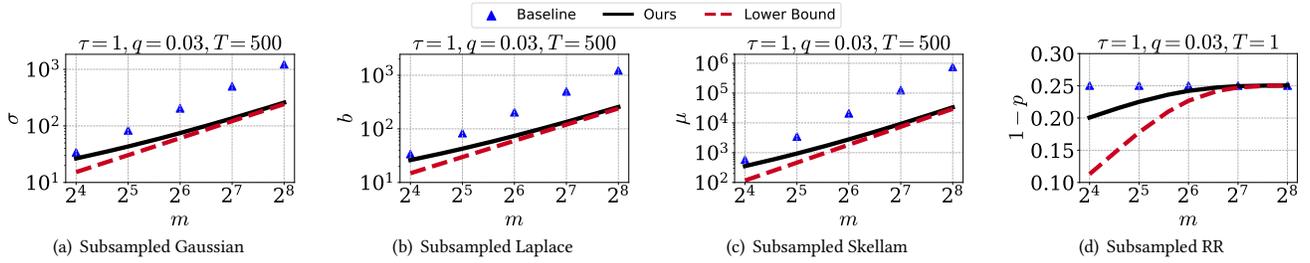

**Figure 2: Varying the group size $m$.**

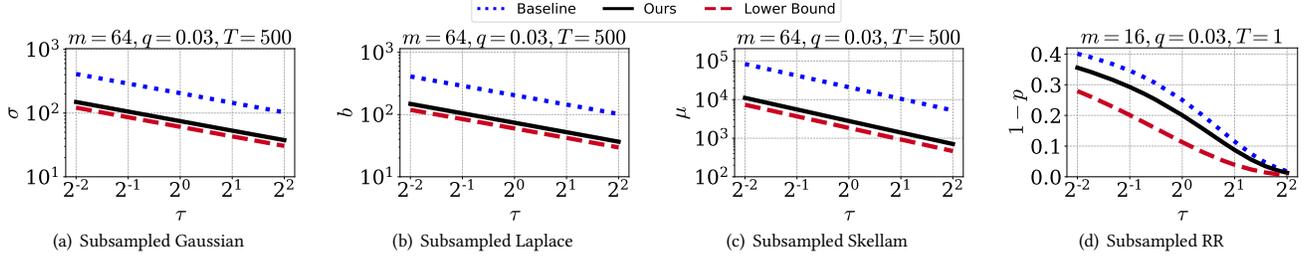

**Figure 3: Varying the privacy parameter $\tau$.**

# 5 EXPERIMENTS

This section presents comparative experiments between our proposed RGP bound and the existing method with generic DP-to-GP conversion. We start by comparing the required noise levels for achieving a given $(m, \alpha, \tau)$-RGP guarantee through different RGP bounds in Section 5.1. Then, we evaluate the practicability of our RGP solution using the DP-SGD algorithm for neural network model training in Section 5.2. Specifically, the RGP bounds examined in our experiments are listed as follows:

(1) The baseline method, which is derived by first obtaining the RDP guarantee of the subsampled mechanism according to the RDP bound in [63], then converting the RDP guarantee to the RGP guarantee using Lemma 2.3.

(2) Our closed-form RGP bounds in Section 4.

In addition, to demonstrate the tightness of our RGP bounds, we compare the noise levels required by our bounds with the noise levels calibrated by the analytical lower bounds across different subsampled mechanisms. The details of the analytical lower bounds can be found in Appendix D.

Note that the closed-form RGP guarantees in Section 4 apply to both unbounded and bounded RGP notions. From the perspective of algorithmic implementation, ensuring unbounded (or bounded) RGP necessitates bounding the difference $\|f(D) - f(\hat{D})\|$ for all pairs of unbounded (or bounded) neighbor datasets $D$ and $\hat{D}$ by a constant $C$. In light of this, we specify the value of $C$ but do not differentiate between unbounded and bounded cases. The comparative results between unbounded and bounded RGP bounds are identical in the experiments.

## 5.1 Comparison of Noise Levels

In this subsection, we conduct a numerical comparison of RGP bounds across different configurations of the Gaussian, Laplace, Skellam, and RR mechanisms. Specifically, for a given $(m, \alpha, \tau)$-RGP guarantee, we calibrate the required noise parameters ($\sigma$ for Gaussian, $b$ for Laplace, $\mu$ for Skellam, and $p$ for the RR mechanism) to achieve this guarantee through our proposed analysis, the baseline solution, and the theoretical lower bound. In the experiments, we set the norm bound $C$ to 1 for the Skellam mechanism; for Gaussian and Laplace mechanisms, the norm bound is not specified as it is canceled out in the final expression of the RGP guarantee.

**Varying the group size $m$.** We first vary the group size from 16 to 256 while keeping other parameters fixed, and then compare the required noise levels derived from our analysis, the baseline method, and the theoretical lower bound. Specifically, we fix the privacy guarantee to $(m, \alpha = 4, \tau = 1)$-RGP, which can be converted to $(m, \epsilon = 4.1, \delta = 10^{-5})$-GP via Lemma 2.2. The results are depicted in Figure 2. Details on the sampling rate $q$, and the number of iterations $T$ are provided in each subfigure. Note that in the RR mechanism, a lower $p$ value indicates larger noise. Therefore, we compare the value of $1 - p$ for the RR mechanism to facilitate a better comparison of noise levels.

Our main observations from the experimental results are threefold: (i) for Gaussian, Laplace, and Skellam mechanisms, our bound is not only tight but also clearly superior to the baseline; (ii) as $m$ increases, the gap between our bound and the baseline method grows for the Gaussian mechanism, aligning with the theoretical insights from Theorem 4.2; (iii) for the RR mechanism, while our bound consistently outperforms the baseline, there is a gap between our bound and the theoretical lower bound at small group sizes, suggesting room for further improvement in RGP guarantees.

**Varying the privacy parameter $\tau$.** We vary the RGP privacy parameter $\tau$ by $\{0.25, 0.5, 1, 2, 4\}$ and calibrate the noise levels to achieve $(m, \alpha = 4, \tau)$-RGP guarantee. These RGP guarantees can be converted (i.e., using Lemma 2.2) to $(\epsilon, \delta = 10^{-5})$-GP guarantees, with the corresponding $\epsilon$ values being 3.4, 3.6, 4.1, 5.2, and 7.1, respectively. The experimental results are presented in Figure 3. From these results, we observe that the required noise levels calibrated by our bounds are consistently and significantly smaller than those required by the baseline method. Furthermore, for Gaussian, Laplace, and Skellam mechanisms, our privacy analysis results tightly match the corresponding theoretical lower bounds.

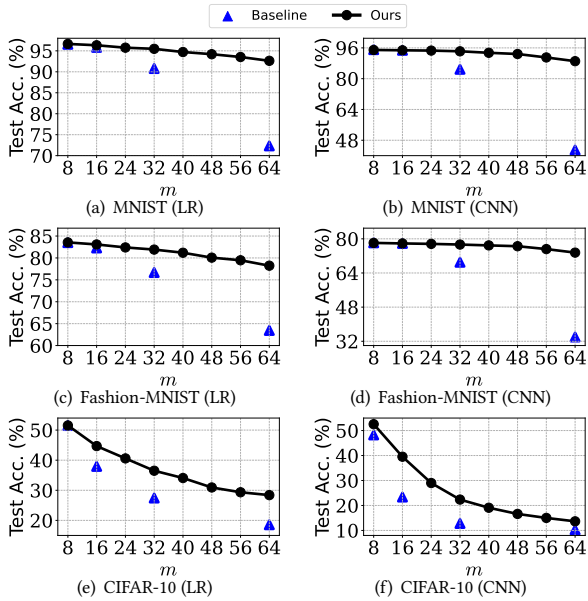

**Figure 4: DP-SGD with $(m, 4, 10^{-5})$-GP guarantee.**

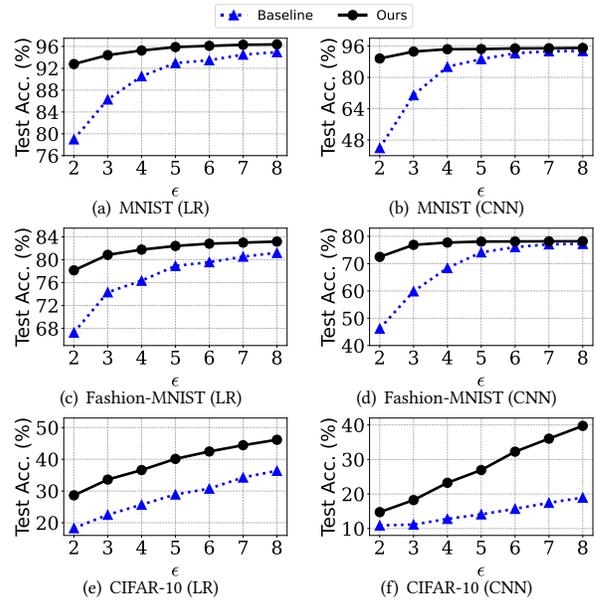

**Figure 5: DP-SGD with $(32, \epsilon, 10^{-5})$-GP guarantee.**

We also compare the RGP bounds by varying the sampling rate and the number of iterations. The experimental results are reported in Appendix F.

**Summary of results.** All above experimental results validate that our RGP bound is consistently tight across a wide range of mechanism configurations. Furthermore, these results show over an order of magnitude improvement of our privacy bound compared to the baseline solution, which enables data privacy practitioners to develop practical mechanisms that offer meaningful RGP guarantees while preserving desirable utility levels. In addition, results of the lower bound indicate that ensuring RGP inevitably requires the injection of noise proportional to the group size, which can significantly impair the utility of the mechanism. Therefore, an interesting direction for future research is to explore new methods to relax and redefine the notion of group privacy, which can help design privacy-preserving mechanisms that offer a more favorable balance between group privacy protection and result utility.

### 5.2 Evaluations on DP-SGD

In this subsection, we conduct experiments on model training via the DP-SGD algorithm [1] with different privacy guarantees to validate the practicability and superiority of our RGP bound. It is important to note that the DP-SGD algorithm essentially operates as a type of subsampled Gaussian mechanism [44]. Therefore, given specific privacy parameters, we can calibrate the required Gaussian noise variance based on our closed-form RGP bounds for subsampled Gaussian mechanisms, as presented in Theorem 4.1. Our experiments involve comparisons between two different implementations of the DP-SGD algorithm: one implementation calibrates the noise to achieve group privacy according to Theorem 4.1, and the other calibrates noise based on the baseline solution.

Note that DP-SGD offers $(m, \epsilon, \delta)$-unbounded GP guarantees [1]. For consistency, we also focus on the $(m, \epsilon, \delta)$-unbounded GP in the following experiments.

**Setup.** We conduct experiments on three benchmark image classification datasets: MNIST [35], Fashion-MNIST [57], and CIFAR-10 [34]. Note that in many practical scenarios, image classification tasks require formal group privacy guarantees. For example, it is necessary to prevent the inference of gender or race ratios within a specific group in human face datasets [39]. Similarly, it is essential to safeguard information about genetic diseases within groups, such as families, in medical image datasets.

Following the state-of-the-art differentially private learning solution [50], we use logistic regression (LR) and convolutional neural networks (CNN) paired with Scattering Networks (SN) in our experiments. We defer the details of the experimental setup to Appendix E. In all experiments, we fix the privacy parameter $\delta$ to $10^{-5}$.

To achieve an $(m, \epsilon, \delta = 10^{-5})$-GP guarantee, we employ a binary search method to determine the required Gaussian noise scale $\sigma$. For each noise scale $\sigma$, we initially obtain $(m, \alpha, \tau)$-RGP guarantees for each $\alpha$ in the set $\{2, 3, \ldots, 100\}$ based on our closed-form RGP bound for subsampled Gaussian mechanism in Theorem 4.1. Subsequently, we calculate the corresponding privacy parameters $\epsilon$ for each $\alpha$ by Lemma 2.2. The minimum $\epsilon$ identified in this process is then selected as the GP guarantee.

**Varying the group size $m$.** In this experiment, we set the group privacy parameter $\epsilon$ to 4 and vary the group size $m$ by $\{8, 16, 24, 32, 40, 48, 56, 64\}$ to show the impact of different group sizes on the model utility. It is important to note that the baseline solution in Lemma 2.3 can only ensure RGP for group sizes that are powers of two. Therefore, we vary the group size $m$ to $\{8, 16, 32, 64\}$ for the baseline solution.

Figure 4 presents the model test accuracy under different group sizes. We observe that for small group sizes, e.g., $m \leq 16$, our solution performs similarly to the baseline solution. However, for larger group sizes such as $m \geq 32$, our solution consistently and significantly outperforms the baseline solution, while maintaining a desirable model utility. In addition, for all experimental results except for CNN on the CIFAR-10 dataset, we see that the model accuracy gap between our solution and the baseline grows as $m$ increases, confirming our theoretical insights from Theorem 4.2 and aligning with the numerical results in Figure 2(a). For the experimental result of CNN on CIFAR-10 in Figure 4(f), with a smaller group size, e.g., when $m \leq 16$, the model accuracy gap between the proposed and the baseline solutions grows with $m$. When $m$ is relatively large (e.g., over 32), however, the gap no longer expands since the noise levels calibrated by the baseline solution become too large for the deep CNN model to converge, leading to model accuracy close to random guessing (i.e., $\approx 10\%$).

**Varying the privacy parameter $\epsilon$.** Figure 5 reports the model test accuracy for a fixed group size of $m = 32$ across different privacy parameters $\epsilon$. We observe that our solution consistently and significantly outperforms the baseline solution for all settings of $\epsilon$. Specifically, while achieving the same level of group privacy guarantee, our solution dramatically reduces the noise injected into the model weights compared to the baseline method, thus producing models with higher utility, which are more suitable for their target real-world applications.

## 6 RELATED WORK

As a natural extension of DP, the notion of group privacy provides formal protection for the aggregate information of a group of individuals and has been widely recognized and explored in recent years [21, 42, 49, 51]. Previous studies [21, 43, 51] have developed general methodologies to convert DP guarantees to GP guarantees. Nonetheless, as explained in Sections 1 and 2.2, these methods treat DP mechanisms as black boxes to ensure general applicability, resulting in overly conservative results. This limitation has been a driving motivation behind this paper.

We note that a concurrent work [48] also studies RGP bounds for subsampled mechanisms. The main differences between our work and this concurrent study are as follows. First, the RGP bound in [48] involves integrals on a rather complicated function, in which both the numerator and denominator are higher powers of the PDFs of mixture distributions. The number of mixture components — and thus the complexity of the integral function — increases with the group size, leading to numerical instability and inaccuracies for large groups. In contrast, our work provides closed-form RGP bounds that can be efficiently computed with numerical tools. Second, the RGP bound in [48] requires non-trivial and *mechanism-specific* theoretical privacy analysis for each mechanism. This complexity makes it challenging to apply to mechanisms with intricate noise distributions, such as the Skellam mechanism. On the contrary, our RGP bound is concise, requiring only the derivation of

RGP guarantees for the corresponding non-subsampled mechanism. As a result, it can be easily applied to derive closed-form RGP bounds for various subsampled mechanisms, as demonstrated in Section 4. Finally, we established analytical lower bounds to empirically verify the tightness of our bounds and provided experimental evaluations on logistic regression and deep neural networks using benchmark datasets, demonstrating the practical significance of our contributions. In contrast, the concurrent work mainly focuses on theoretical analysis and does not provide any empirical evaluations.

In addition to group privacy, there have been several studies on the protection of user-level privacy in recent years. Particularly, user-level DP [4, 12, 23, 24, 36, 38, 41] has garnered considerable attention. Various practical algorithms ensuring user-level DP have been proposed in [4, 23, 41]. Meanwhile, some studies [12, 24, 38] primarily focus on the theoretical aspects. Notable among these are recent works [12, 24], which have delved into methods of converting DP guarantees into user-level DP guarantees. It is crucial to note, however, that these studies rely on asymptotic analysis with additional assumptions, such as i.i.d. distribution of data records. Consequently, while these works contribute valuable theoretical insights, they fall short of providing precise, closed-form privacy guarantees that are essential for designing practical privacy-preserving mechanisms. Furthermore, while user-level DP does offer a degree of group-level privacy, its definition is fundamentally distinct from that of group privacy. Specifically, user-level DP assumes that data is collected from various users, each holding $m$ records [23, 24, 38]. Under this paradigm, user-level DP offers theoretical privacy protections for records associated with any specific user. In contrast, group privacy safeguards information about any group of $m$ records, thereby inherently encompassing and extending beyond the scope of user-level DP. Therefore, group privacy is more general than user-level DP, and thus algorithms satisfying user-level DP (such as those in [36, 41]) may not ensure group privacy.

## 7 CONCLUSION

In this paper, we present a tight general RGP bound applicable to subsampled mechanisms, based on which we derive precise and closed-form group privacy guarantees for a variety of prevalent privacy-preserving mechanisms, including subsampled Gaussian, Laplace, Skellam, and Randomized Response. For the $d$-dimensional Laplace mechanism within the RDP framework, our refined analysis yields a significantly tighter RDP bound, which offers a multiplicative factor saving of $d$ and may be of independent interest. Experimental results demonstrate both the tightness and a substantial improvement of our bound over existing RGP guarantees. To the best of our knowledge, this is the first work that presents tight and closed-form RGP guarantees for subsampled mechanisms.

Regarding future work, we plan to further refine the RGP bound for subsampled Randomized Response mechanisms, and to derive closed-form RGP guarantees for other types of subsampled mechanisms such as DPIS [55], in which each record has a different probability of being included in the sample set.

## A   KEY INSIGHTS AND INTUITIONS

Below, we elaborate on the key insights and intuitions behind our theoretical results on the group privacy guarantees for Poisson subsampled mechanisms.

### A.1   Reducing Group Influence

By applying Poisson sampling, one can reduce the influence of any specific group on the output of the mechanism; thus, only a small noise magnitude is required to mask the presence of the group. Let $X$ denote the number of sampled records in a group of size $m$ and let $q$ denote the sampling rate. Then $X$ can be viewed as a random variable that follows the binomial distribution:

$$\Pr[X = k] = \binom{m}{k}(1 - q)^{(m-k)}q^k.$$

Note that the above binomial distribution has mean $mq$ and variance $mq(1-q)$. Therefore, for *any group* of size $m$, when $q$ is small, the number of sampled records in the group would be tightly concentrated around $mq$ instead of $m$. For example, when $m = 100$ and $q = 0.01$, by applying Chebyshev's inequality, we can verify that

$$\Pr[|X - 1| \geq 4] = \Pr[|X - \mathbb{E}[X]| \geq 4] \leq \frac{\text{Var}[X]}{4^2} = \frac{0.99}{16} < 0.07.$$

In other words, for *any specific group* with size $m = 100$, with a high probability of at least 93%, there will be at most 5 group members being sampled into the batch. This implies that the group-level privacy leakage (i.e., the presence of any specific group in the dataset) about *any group* is caused by 5 records instead of the entire group members (i.e., 100 records) in most cases.

More generally, when $q$ is small, e.g., $qm = O(1)$, with high probability, the potential information leakage about *any group* of size $m$ is mainly caused by $O(qm)$ records, hence the noise roughly proportional to $qm$ instead of $m$ would be enough for protecting group privacy.

### A.2   Group-level Sampling

Poisson sampling can be viewed as "group-level" sampling — it includes each group in the batch with a certain probability. Specifically, given *any group* of size $m$, let $q \in (0, 1)$ be the Poisson sampling rate. The probability that none of the group members are sampled into the batch is

$$\Pr[\text{no record in the group is sampled}]$$
$$= (1 - q)^m = \left(1 - \frac{qm}{m}\right)^m \geq e^{-qm} \cdot (1 - q^2 m),$$

where the inequality follows from the fact that $(1 + \frac{x}{n})^n \geq e^x(1 - \frac{x^2}{n})$ for $n \geq 1, |x| \leq n$. The above inequality implies that when the sampling rate $q$ is small enough such that $qm = O(1)$, for *any specific group*, none of the group members will not be sampled into the batch with high probability. For example, let $m = 10$ and $q = 0.01$; then any group with size $m = 10$ will not be sampled with a high probability of at least $e^{-0.1}(1 - 0.001) > 90\%$. In other words,

with a probability of at least 90%, there will be no group privacy leakage for any specific group of size $m = 10$.

## B   USEFUL LEMMAS

The following lemmas are used in Section 4 to derive closed-form group privacy guarantees for various subsampled mechanisms..

**Lemma B.1** (Quasi-convexity of Rényi Divergence [52]). *For any two pairs of probability distributions $(P, Q)$ and $(P', Q')$, and any $\beta \in (0, 1)$, we have*

$$D_\alpha\big((1 - \beta)P + \beta P' \,\big\|\, (1 - \beta)Q + \beta Q'\big)$$
$$\leq \max\big\{D_\alpha(P\|Q), D_\alpha(P'\|Q')\big\}.$$

Note that the quasi-convexity lemma presented in [52] is with two mixture components. By induction, it can be verified that the above quasi-convexity lemma extends to multiple mixture components as follows.

**Corollary B.1.** *For any $k$ probability distributions $(Q_1, \ldots, Q_k)$ and $(Q'_1, \ldots, Q'_k)$, and any set of coefficients of convex combination $\{w_1, w_2, \ldots, w_k\}$ such that $\sum_{i=1}^k w_i = 1$ and $w_i \geq 0$ for all $i \in \{1, 2, \ldots, k\}$, we have*

$$D_\alpha\left(\sum_{i=1}^k w_i Q_i \,\bigg\|\, \sum_{i=1}^k w_i Q'_i\right) \leq \max_{i \in \{1,2,\ldots,k\}}\big\{D_\alpha(Q_i\|Q'_i)\big\}.$$

PROOF. The corollary is established by applying mathematical induction to Lemma B.1. □

**Lemma B.2** (Convexity in the Second Term [52]). *For any probability distributions $P$, $Q$, and $Q'$, it holds that*

$$D_\alpha(P\|(1 - \beta)Q + \beta Q') \leq (1 - \beta)D_\alpha(P\|Q) + \beta D_\alpha(P\|Q'),$$

*for any $\beta \in (0, 1)$.*

**Lemma B.3** (Additivity of Rényi Divergence [52]). *For any $d \in \mathbb{N}$ and arbitrary distributions $P_1, P_2, \ldots, P_d$ and $Q_1, Q_2, \ldots, Q_d$, let $P^d := P_1 \times \cdots \times P_d$ and $Q^d := Q_1 \times \cdots \times Q_d$. Then*

$$D_\alpha(P^d\|Q^d) = \sum_{i=1}^d D_\alpha(P_i\|Q_i).$$

**Lemma B.4** (Joint Convexity [53]). *For all $\alpha > 1$, the binary function $f(x, y) = x^\alpha/y^{\alpha-1}$ is jointly convex on $\mathbb{R}_+^2$.*

The following lemmas present the closed-form Rényi divergence for the distributions of Gaussian, Laplace, Skellam, and the mechanism of Randomized Response.

**Lemma B.5** ([25]). *Let $\mathcal{N}(\mu, \sigma^2)$ denote the one-dimensional Gaussian distribution with mean $\mu$ and variance $\sigma^2$, it holds that*

$$D_\alpha\big(\mathcal{N}(\mu, \sigma^2)\big\|\mathcal{N}(0, \sigma^2)\big) = D_\alpha\big(\mathcal{N}(0, \sigma^2)\big\|\mathcal{N}(\mu, \sigma^2)\big)$$
$$= \frac{\alpha\mu^2}{2\sigma^2}.$$

**Lemma B.6** ([25]). *Let $\text{Lap}(\mu, b)$ denote the one-dimensional Laplace distribution with mean $\mu$ and scale $b$, we have*

$$D_\alpha(\text{Lap}(\mu, b)\|\text{Lap}(0, b)) = D_\alpha(\text{Lap}(0, b)\|\text{Lap}(\mu, b))$$
$$= \frac{1}{\alpha - 1}\log\left\{\frac{\alpha}{2\alpha - 1}\exp\left(\frac{(\alpha - 1)\mu}{b}\right) + \frac{\alpha - 1}{2\alpha - 1}\exp\left(\frac{-\alpha\mu}{b}\right)\right\}.$$

**Lemma B.7** ([3]). *Let* $\text{Sk}(z, \mu)$ *denote the symmetric Skellam distribution with mean* $z \in \mathbb{Z}$ *and variance* $\mu$, *we have*

$$D_{\alpha}(\text{Sk}(z, \mu) \| \text{Sk}(0, \mu)) \leq \frac{\alpha z^2}{2\mu} + \min\left\{\frac{(2\alpha - 1)z^2 + 6|z|}{4\mu^2}, \frac{3|z|}{2\mu}\right\}.$$

**Lemma B.8** ([43]). *Let* $\mathcal{A}$ *be the randomized response mechanism defined in* (19). *For any pair of dataset* $D$ *and* $D'$, *it holds that*

$$D_{\alpha}(\mathcal{A}(D) \| \mathcal{A}(D')) \leq \frac{1}{\alpha - 1} \log\left(\frac{p^{\alpha}}{(1 - p)^{\alpha - 1}} + \frac{(1 - p)^{\alpha}}{p^{\alpha - 1}}\right).$$

# C PROOFS

## C.1 Proof of Theorem 4.2

**Theorem 4.2.** *Consider a subsampled Gaussian mechanism* $\mathcal{M}$ *with a sampling rate* $q \in (0, 1)$ *and variance* $\sigma^2$. *Let* $\alpha > 1$ *be any integer. We denote by* $(m, \alpha, \tau'_m(\alpha))$ *and* $(m, \alpha, \tau_m(\alpha))$ *the unbounded RGP guarantees of* $\mathcal{M}$ *derived using Lemma 2.3 and Theorem 4.1, respectively. If* $\sigma = \Theta(m)$ *and* $qm > 1$, *our bound* $\tau_m(\alpha)$ *saves a multiplicative factor of* $\Theta(m^{\log_2 1.5})$ *compared to* $\tau'_m(\alpha)$.

PROOF. By Theorem 4.1 in Section 4, the proposed Rényi group privacy guarantee for the subsampled Gaussian mechanism with group size $m$ can be derived as:

$$\tau_m(\alpha) = \frac{1}{\alpha - 1} \log\underbrace{\left(\sum_{k=0}^{m} \binom{m}{k}(1 - q)^{m-k} q^k \exp\left(\frac{(\alpha - 1)\alpha k^2}{2\sigma^2}\right)\right)}_{G_m(\alpha)}.$$

To obtain $\tau'_m(\alpha)$, we need to convert RDP guarantee for the subsampled Gaussian mechanism to the Rényi group privacy via Lemma 2.3. The following lemma presents the RDP guarantee for the subsampled Gaussian mechanism:

**Lemma C.1** (RDP for the Subsampled Gaussian Mechanism [44]). *The subsampled Gaussian mechanism with sampling rate* $q$ *and variance* $\sigma^2$ *satisfies* $(\alpha, \tau(\alpha))$-*unbounded RDP, where*

$$\tau(\alpha) = \frac{1}{\alpha - 1} \log\left(\sum_{i=0}^{\alpha} \binom{\alpha}{i}(1 - q)^{\alpha - i} q^i \exp\left(\frac{i^2 - i}{2\sigma^2}\right)\right).$$

Then by Lemma 2.3 and the identity $3^{\log m} = m^{\log 3}$, we can derive that $\mathcal{M}$ satisfies $(m, \alpha, \tau'_m(\alpha))$-unbounded Rényi group privacy, where

$$\tau'_m(\alpha) = \frac{m^{\log_2 3}}{\alpha m - 1} \log\underbrace{\left(\sum_{i=0}^{\alpha m} \binom{\alpha m}{i}(1 - q)^{\alpha m - i} q^i \exp\left(\frac{i^2 - i}{2\sigma^2}\right)\right)}_{H_m(\alpha)}.$$

In what follows, we first bound the terms $G_m(\alpha)$ and $H_m(\alpha)$, and then compare $\tau_m(\alpha)$ and $\tau'_m(\alpha)$ as a whole. Before proceeding to the analysis, we first introduce a useful lemma.

**Lemma C.2.** *Let* $X \sim B(m, q)$, *where* $q \in (0, 1)$ *and* $m \in \mathbb{N}$. *Suppose* $qm > 1$, *then for* $t \geq 1$, *there is a constant* $C_t$ *such that*

$$(qm)^t < \mathbb{E}[X^t] \leq C_t (qm)^t.$$

The proof of Lemma C.2 follows from Theorem 4.1 in [32] and Lemma 27 in [2], and is omitted in this paper for clarity of presentation. If $\sigma = \Theta(m)$, it can be easily verified by Taylor's theorem with Lagrange remainder that there exists a constant $N_0 \in \mathbb{N}$ such that for all $N \geq N_0$, we have

$$\exp\left(\frac{(\alpha - 1)\alpha k^2}{2\sigma^2}\right) = \sum_{n=0}^{N} \frac{1}{n!}\left(\frac{(\alpha - 1)\alpha k^2}{2\sigma^2}\right)^n + r_1 \quad (\forall k \in [m]) \quad (21)$$

and

$$\exp\left(\frac{i^2 - i}{2\sigma^2}\right) = \sum_{n=0}^{N} \frac{1}{n!}\left(\frac{i^2 - i}{2\sigma^2}\right)^n + r_2 \quad (\forall i \in [\alpha m]), \quad (22)$$

where $r_1, r_2 \in (0, 1]$. Let $X$ denote the random variable draw from the binomial distribution $B(m, q)$, then by Lemma C.2 and (21), we have

$$
\begin{aligned}
G_m(\alpha) &= \sum_{k=0}^{m} \binom{m}{k}(1 - q)^{m-k} q^k \left(\sum_{n=0}^{N} \frac{1}{n!}\left(\frac{(\alpha - 1)\alpha k^2}{2\sigma^2}\right)^n + r_1\right) \\
&\leq \sum_{n=0}^{N} \frac{1}{n!} \sum_{k=0}^{m} \binom{m}{k}(1 - q)^{m-k} q^k \left(\frac{(\alpha - 1)\alpha k^2}{2\sigma^2}\right)^n + 1 \\
&= \sum_{n=0}^{N} \frac{1}{n!}\left(\frac{(\alpha - 1)\alpha}{2\sigma^2}\right)^n \mathbb{E}[X^{2n}] + 1 \\
&\leq \sum_{n=0}^{N} \frac{1}{n!}\left(\frac{(\alpha - 1)\alpha q^2}{2}\right)^n \left(\frac{m}{\sigma}\right)^{2n} C_{2n} + 1 \\
&= \sum_{n=0}^{N} \frac{1}{n!}\left(\frac{(\alpha - 1)\alpha q^2}{2}\right)^n \Theta(1) \\
&= e^{\Theta((\alpha - 1)\alpha q^2 / 2)},
\end{aligned}
$$

where the last equation holds when $N$ is sufficiently large.

Let $Y$ denote the random variable draw from the binomial distribution $B(\alpha m, q)$, then by a similar analysis, the term $H_m(\alpha)$ can be lower bounded as follows:

$$
\begin{aligned}
H_m(\alpha) &= \sum_{i=0}^{\alpha m} \binom{\alpha m}{i}(1 - q)^{\alpha m - i} q^i \left(\sum_{n=0}^{N} \frac{1}{n!}\left(\frac{i^2 - i}{2\sigma^2}\right)^n + r_2\right) \\
&\geq \sum_{n=0}^{N} \frac{1}{n!} \sum_{i=0}^{\alpha m} \binom{\alpha m}{i}(1 - q)^{\alpha m - i} q^i \left(\frac{i^2 - i}{2\sigma^2}\right)^n \\
&\geq \sum_{n=0}^{N} \frac{1}{n!}\left(\frac{1}{2\sigma^2}\right)^n \left[\sum_{i=0}^{\alpha m} \binom{\alpha m}{i}(1 - q)^{\alpha m - i} q^i \left(\frac{i^2}{2}\right)^n \right. \\
&\qquad\qquad\qquad\left. - \binom{\alpha m}{1}(1 - q)^{\alpha m - 1} q\left(\frac{1}{2}\right)^n\right] \\
&= \sum_{n=0}^{N} \frac{1}{n!}\left(\frac{1}{4\sigma^2}\right)^n \left[\mathbb{E}[Y^{2n}] - \binom{\alpha m}{1}(1 - q)^{\alpha m - 1} q\right] \\
&\geq \sum_{n=0}^{N} \frac{1}{n!}\left(\frac{\alpha^2 q^2}{4\sigma^2}\right)^n m^{2n} - \alpha m(1 - q)^{\alpha m - 1} q e^{1/(4\sigma^2)} \\
&= \sum_{n=0}^{N} \frac{1}{n!}\left(\frac{\alpha^2 q^2}{4}\right)^n \Theta(1) - \alpha m(1 - q)^{\alpha m - 1} q e^{1/(4\sigma^2)} \\
&= e^{\Theta(\alpha^2 q^2 / 4)} - \alpha m(1 - q)^{\alpha m - 1} q e^{1/(4\sigma^2)}
\end{aligned}
$$

where the first inequality follows from $r_2 > 0$; the second inequality follows from $i^2 - i \geq i^2/2$ for all $i \in \mathbb{N} \setminus \{1\}$; the third inequality follows from Lemma C.2.

Note that the term $\alpha m(1-q)^{\alpha m-1}qe^{1/(4\sigma^2)}$ goes to 0 as $m \to \infty$, which implies that $H_m(\alpha)$ is dominant by the term $e^{\Theta(\alpha^2 q^2/4)}$ as $m$ goes to large. Therefore, we have

$$\tau_m(\alpha) = \frac{1}{\alpha-1}\log G_m(\alpha) \leq \frac{1}{\alpha-1}\Theta((\alpha-1)\alpha q^2) = \Theta(\alpha q^2)$$

and

$$\tau'_m(\alpha) = \frac{m^{\log_2 3}}{\alpha m-1}\log H_m(\alpha) \geq \frac{m^{\log_2 3}}{\alpha m-1}\Theta(\alpha^2 q^2) = \Theta(m^{\log_2 1.5}\alpha q^2).$$

The theorem follows. □

## C.2 Proof of Lemma 4.3

**Lemma 4.3.** *The function $f(x) = \log\left(c_1 e^{\beta_1 x} + c_2 e^{-\beta_2 x}\right)$ is convex for $c_1, c_2, \beta_1, \beta_2 \in (0, \infty)$.*

Proof. By definition of convexity, it suffices to show that $f''(x) \geq 0$. First, note that

$$f'(x) = \frac{\beta_1 c_1 e^{\beta_1 x} - \beta_2 c_2 e^{-\beta_2 x}}{c_1 e^{\beta_1 x} + c_2 e^{-\beta_2 x}}.$$

Let $h(x) = \beta_1 c_1 e^{\beta_1 x} - \beta_2 c_2 e^{-\beta_2 x}$ and $g(x) = c_1 e^{\beta_1 x} + c_2 e^{-\beta_2 x}$, then we have

$$f''(x) = \left(\frac{h(x)}{g(x)}\right)' = \frac{h'(x)g(x) - h(x)g'(x)}{(g(x))^2}.$$

Note that $g(x) > 0$ for all $x \in \mathbb{R}$. Hence to show $f''(x) \geq 0$, it suffices to verify that $h'(x)g(x) - h(x)g'(x) \geq 0$. The first derivative of $h(x)$ and $g(x)$ is as follows

$$h'(x) = \beta_1^2 c_1 e^{\beta_1 x} + \beta_2^2 c_2 e^{-\beta_2 x},$$
$$g'(x) = \beta_1 c_1 e^{\beta_1 x} - \beta_2 c_2 e^{-\beta_2 x}.$$

Then by a direct calculation, we obtain

$$h'(x)g(x) - h(x)g'(x)$$
$$= \beta_1^2 c_1 c_2 e^{(\beta_1-\beta_2)x} + \beta_2^2 c_1 c_2 e^{(\beta_1-\beta_2)x} + 2\beta_1 \beta_2 c_1 c_2 e^{(\beta_1-\beta_2)x} \geq 0.$$

The lemma is proved. □

## C.3 Proof of Lemma 4.5

**Lemma 4.5.** *For the RR mechanism, it holds that $\lim_{m\to\infty} \tau_m(\alpha) = \tau_1^*(\alpha)$. In addition, we have $\tau_m(\alpha) < \tau_1^*(\alpha)$ for all $m \in \mathbb{N}$.*

Proof. Note that $\lim_{m\to\infty}(1-q)^m = 0$, which immediately yields the first claim of Lemma 4.5. To prove the second claim, we need the following lemma.

**Lemma C.3.** *Let $\varphi(p) = \frac{p^\alpha}{(1-p)^{\alpha-1}} + \frac{(1-p)^\alpha}{p^{\alpha-1}}$. For all $\alpha \geq 2$ and $p \in (0.5, 1)$, we have $\varphi(p) > 1$.*

The proof of Lemma C.3 is deferred to the end of this section. Let $\Phi^{RR}(\alpha) := \frac{p^\alpha}{(1-p)^{\alpha-1}} + \frac{(1-p)^\alpha}{p^{\alpha-1}}$. By Lemma C.3, it holds that $1 < \Phi^{RR}(\alpha)$. Therefore, we have

$$\tau_m(\alpha) = \frac{1}{\alpha-1}\log\left((1-q)^m \cdot 1 + (1-(1-q)^m) \cdot \Phi^{RR}(\alpha)\right)$$
$$< \frac{1}{\alpha-1}\log\left((1-q)^m \cdot \Phi^{RR}(\alpha) + (1-(1-q)^m) \cdot \Phi^{RR}(\alpha)\right)$$
$$= \frac{1}{\alpha-1}\log \Phi^{RR}(\alpha)$$
$$= \tau_1^*(\alpha).$$

Lemma 4.5 is proved. □

Proof of Lemma C.3. The claim in Lemma C.3 can be verified by examining the monotonicity of function $\varphi(p)$. The first-order derivative of $\varphi(p)$ is

$$\varphi'(p) = \frac{p^{\alpha-1}(\alpha-p)}{(1-p)^\alpha} + \frac{(1-p)^{\alpha-1}(1-\alpha-p)}{p^\alpha}$$
$$= \frac{p^{2\alpha-1}(\alpha-p) + (1-p)^{2\alpha-1}(1-\alpha-p)}{p^\alpha(1-p)^\alpha}. \quad (23)$$

Since $p \in (0.5, 1)$ and $\alpha \geq 2$, the denominator of (23) is strictly greater than 0. Hence, the sign of $\varphi'(p)$ is determined by its numerator. Let $h_1(p) = p^{2\alpha-1}(\alpha-p)$ and $h_2(p) = (1-p)^{2\alpha-1}(1-\alpha-p)$. We now examine the monotonicity of $h_1$ and $h_2$. For $h_1$, we have

$$h_1'(p) = p^{2\alpha-2}(\alpha(2\alpha-1) - 2\alpha p)$$
$$> p^{2\alpha-2}(2\alpha^2 - 3\alpha)$$
$$> 0.$$

For $h_2$, we have

$$h_2'(p) = (1-2\alpha)(1-p)^{2\alpha-2}(1-\alpha-p) - (1-p)^{2\alpha-1}$$
$$= (1-p)^{2\alpha-2}\left(2\alpha^2 + 2\alpha p - 3\alpha\right)$$
$$> (1-p)^{2\alpha-2}\left(2\alpha^2 - 3\alpha\right)$$
$$> 0.$$

Therefore, the function $h(p) = h_1(p) + h_2(p)$ is a strictly monotonically increasing function for $p \in (0.5, 1)$. Observing that $h(0.5) = 0$ and considering the strict monotonicity of $h$, we conclude that $h(p) > 0$. Since the sign of $\varphi'(p)$ is identical to $h(p)$, we can conclude that $\varphi(p)$ is strictly increasing on $(0.5, 1)$. Hence, we have $\varphi(p) > \varphi(0.5) = 1$ for all $p \in (0.5, 1)$. The lemma is proved. □

# D ANALYTICAL LOWER BOUNDS

In this section, we establish the analytical lower bounds for the subsampled mechanisms discussed in Section 4. Note that these lower bounds are used in the numerical experiments of Section 5.1 for empirically validating the tightness of our group privacy upper bounds. Consider a pair of binary datasets $D$ and $D'$ defined as follows:

$$D = \{0, 0, \ldots, 0\} \quad \text{and} \quad D' = D \cup \underbrace{\{1, 1, \ldots, 1\}}_{m \text{ records}}.$$

We now extend our lower bound analysis in Section 3.3 to all subsampled additive mechanisms of the form $\mathcal{M} := \mathcal{A} \circ \text{Subsample}$, where $\mathcal{A}(D) = f(D) + Z$, and $Z$ is a random variable following a specified distribution such as one-dimensional Gaussian, Laplace, and Skellam. Specifically, for a random variable $k + Z$ with $k \in \mathbb{N}$, let $\mu_k$ be its probability density function (pdf). The distributions of $\mathcal{M}(D)$ and $\mathcal{M}(D')$ are then given by:

$$\mathcal{M}(D) \sim \mu_0 \quad \text{and} \quad \mathcal{M}(D') = \sum_{k=0}^m p_k \mu_k,$$

**Table 1: CNN model for MNIST and Fashion-MNIST datasets, with Tanh activations.**

| Layer | Parameters |
|---|---|
| Convolution | 16 filters of $3 \times 3$, stride 2, padding 1 |
| Convolution | 32 filters of $3 \times 3$, stride 1, padding 1 |
| Fully connected | 32 units |
| Fully connected | 10 units |

**Table 2: CNN model for CIFAR-10 dataset, with Tanh activations.**

| Layer | Parameters |
|---|---|
| Convolution | 64 filters of $3 \times 3$, stride 1, padding 1 |
| Max-Pooling | $2 \times 2$, stride 2 |
| Convolution | 64 filters of $3 \times 3$, stride 1, padding 1 |
| Fully connected | 10 units |

**Table 3: Number of learnable parameters.**

| Model | Dataset | Number of parameters |
|---|---|---|
| SN+LR | MNIST/Fashion-MNIST | 39,700 |
| | CIFAR-10 | 155,530 |
| SN+CNN | MNIST/Fashion-MNIST | 33,066 |
| | CIFAR-10 | 187,146 |

where $p_k = \binom{m}{k}(1-q)^{m-k}q^k$. The lower bound is derived as follows:

$$\tau_m(\alpha) \geq D_\alpha(\mathcal{M}(D')\|\mathcal{M}(D))$$
$$= \frac{1}{\alpha - 1} \log \mathbb{E}_{\mu_0} \left[ \left( \frac{\sum_{k=0}^{m} p_k \mu_k}{\mu_0} \right)^\alpha \right].$$

Substituting the pdfs of Gaussian, Laplace, and Skellam distributions into $\mu_k$ establishes the analytical lower bounds for the subsampled mechanisms based on these distributions. It is worth noting that although the resulting lower bounds are not in closed form and require computational effort to determine, their formulations are straightforward and one-dimensional, allowing for efficient and precise computation using numerical tool-kits such as SciPy (https://scipy.org/).

For subsampled RR mechanisms, we examine the task of privately releasing the output of the indicator function $f(D) = \mathbf{1}(\sum_{x \in D} = 0)$, which determines whether all records in a binary dataset $D$ are zeros. Let $\mathcal{A}$ be the RR mechanism defined in Section 4.4. Note that for any binary dataset $\tilde{D}$ containing at least one "1", the pdf of $\mathcal{A}(\tilde{D})$ remains the same. Now, let $\mu_0$ and $\mu_1$ denote the pdfs of $\mathcal{A}(D)$ and $\mathcal{A}(\tilde{D})$, respectively. The distributions of $\mathcal{M}(D)$ and $\mathcal{M}(D')$ are then:

$$\mathcal{M}(D) \sim \mu_0 \quad \text{and} \quad \mathcal{M}(D') \sim (1-q)^m \mu_0 + (1 - (1-q)^m)\mu_1.$$

Hence, the lower bound is:

$$\tau_m(\alpha) \geq D_\alpha(\mathcal{M}(D')\|\mathcal{M}(D))$$
$$= \frac{1}{\alpha - 1} \log \mathbb{E}_{\mu_0} \left[ \left( (1-q)^m + (1 - (1-q)^m)\frac{\mu_1}{\mu_0} \right)^\alpha \right].$$

Given the output distribution of $\mathcal{A}$ defined in (19), this lower bound can be precisely and efficiently computed.

# E EXPERIMENTAL SETUP

## E.1 Model Architectures

We use logistic regression (LR) models and convolutional neural network (CNN) models in DP-SGD experiments. In addition, all models are paired with a Scattering Network (SN) [45], a non-learned feature extractor, following the state-of-the-art differentially private learning solution in [50]. The SN extracts feature vectors of size $81 \times 7 \times 7$ for both the MNIST and Fashion-MNIST datasets, and of size $243 \times 8 \times 8$ for the CIFAR-10 dataset, adhering to the experimental setup in [50].

The architectures of the CNNs for MNIST/Fashion-MNIST and CIFAR-10 are reported in Table 1 and Table 2, respectively. Note that the CNN architectures in Table 1 and Table 2 are based on the source code from [50] (available at https://github.com/ftramer/Handcrafted-DP/blob/main/models.py) and differ slightly from the descriptions provided in [50]. In addition, the number of learnable parameters for each model is reported in Table 3.

## E.2 Hyper-parameter Settings

In all experiments, we set the number of gradient descent iterations $T$ to 500 and the clipping bound $C$ to 0.1. For the CNN on CIFAR-10 dataset, we use a learning rate of 0.5 and a sampling rate $q$ of 0.1. For all other experiments, we set the learning rate to 0.2 and the sampling rate $q$ to 0.05.

# F ADDITIONAL EXPERIMENTS

We conducted additional numerical comparisons of RGP bounds by varying the sampling rate and the number of iterations for the Gaussian, Laplace, Skellam, and RR mechanisms, respectively.

**Varying the sampling rate $q$.** We investigate the impact of varying sampling rates $q$ on the required noise levels by adjusting $q$ from 0.01 to 0.15. Specifically, for each sampling rate $q$, we calibrate the required noise to ensure a $(m, \alpha = 4, \tau)$-RGP guarantee. The results of these calibrations are reported in Figure 6, with details on the RGP privacy parameters provided in each subfigure. The results show that across different mechanisms and sampling rates, our RGP bound remains tight and consistently outperforms the baseline method.

**Varying the number of iterations $T$.** Figure 7 presents the required noise levels across different iteration numbers. In line with our prior experimental results, the noise levels calibrated using our bounds are both tight and substantially improved compared to those calibrated via the baseline method, demonstrating an order-of-magnitude improvement. This underscores the effectiveness of our approach for subsampled and iterative mechanisms, such as the DP-SGD algorithm and its variants.

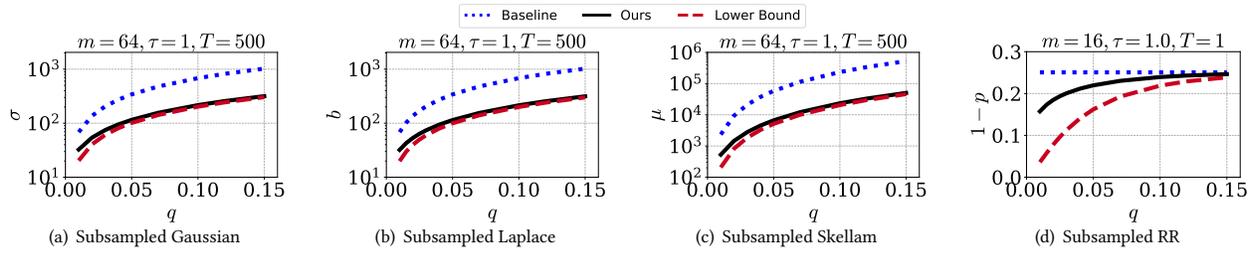

**Figure 6: Varying the sampling rate $q$.**

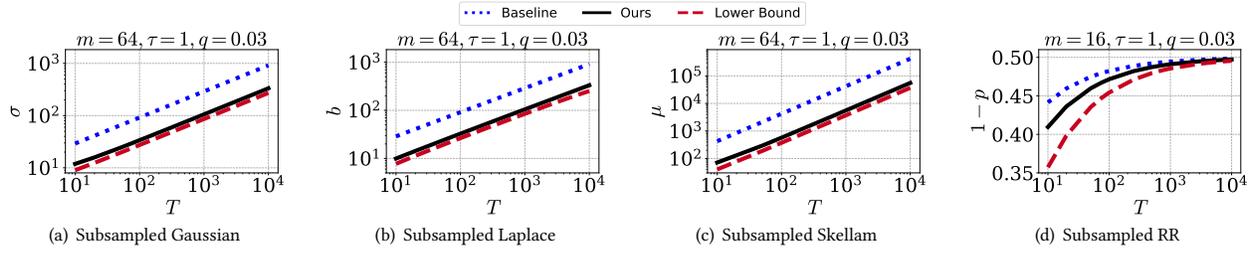

**Figure 7: Varying the number of iterations $T$.**